\documentclass[conference,compsoc]{IEEEtran}

\usepackage{enumitem}
\usepackage{bm}
\usepackage{threeparttable}
\usepackage{booktabs}
\usepackage[framemethod=TikZ]{mdframed}
\usepackage{multirow}
\usepackage{amsmath}
\usepackage{amsfonts}
\usepackage{subcaption}
\usepackage[hidelinks]{hyperref}
\usepackage{tikz}
\newcommand{\pie}[1]{%
\begin{tikzpicture}
 \draw (0ex,0ex) circle (1ex);
 \fill (0ex,-1ex) arc (-90:(#1-90):1ex) -- (0ex,-1ex) -- cycle;
\end{tikzpicture}%
}

\ifCLASSOPTIONcompsoc
  \usepackage[nocompress]{cite}
\else
  \usepackage{cite}
\fi

\ifCLASSINFOpdf

\else

\fi

\begin{document}

\title{Learn What You Want to Unlearn: \\
Unlearning Inversion Attacks against Machine Unlearning}

\author{\IEEEauthorblockN{Hongsheng Hu\IEEEauthorrefmark{1},
Shuo Wang\IEEEauthorrefmark{2}, Tian Dong\IEEEauthorrefmark{2} and Minhui Xue\IEEEauthorrefmark{1} 
}
\IEEEauthorblockA{\IEEEauthorrefmark{1}CSIRO's Data61, Australia}
\IEEEauthorblockA{\IEEEauthorrefmark{2}Shanghai Jiao Tong University, China}}

\maketitle
\thispagestyle{plain}
\pagestyle{plain}

\begin{abstract}
Machine unlearning has become a promising solution for fulfilling the ``right to be forgotten'', under which individuals can request the deletion of their data from machine learning models. However, existing studies of machine unlearning mainly focus on the efficacy and efficiency of unlearning methods, while neglecting the investigation of the privacy vulnerability during the unlearning process. With two versions of a model available to an adversary, that is, the original model and the unlearned model, machine unlearning opens up a new attack surface. 
In this paper, we conduct the first investigation to understand the extent to which machine unlearning can leak the confidential content of the unlearned data. 
Specifically, under the Machine Learning as a Service setting, we propose unlearning inversion attacks that can reveal the feature and label information of an unlearned sample by only accessing the original and unlearned model. 
The effectiveness of the proposed unlearning inversion attacks is evaluated through extensive experiments on benchmark datasets across various model architectures and on both exact and approximate representative unlearning approaches. 
The experimental results indicate that the proposed attack can reveal the sensitive information of the unlearned data. 
As such, we identify three possible defenses that help to mitigate the proposed attacks, while at the cost of reducing the utility of the unlearned model. The study in this paper uncovers an underexplored gap between machine unlearning and the privacy of unlearned data, highlighting the need for the careful design of mechanisms for implementing unlearning without leaking the information of the unlearned data.
\end{abstract}

\IEEEpeerreviewmaketitle

\section{Introduction}\label{sec:introduction}
Data privacy is of paramount importance in today's interconnected digital landscape~\cite{de2021critical}. 
With the increasing volume of personal data being collected, processed, and shared online, the need to protect this sensitive information is critical. 
In response to these concerns, regulations like the General Data Protection Regulation (GDPR)~\cite{rosen2011right} in the European Union and the California Consumer Privacy Act (CCPA)~\cite{pardau2018california} in the United States have been established. 
These regulations aim to empower users by granting them full control over their data, including the ``right to be forgotten'', which allows individuals to request the deletion or removal of their personal data collected and stored by various organizations and services. 
Deleting data from a database can be relatively straightforward~\cite{cao2015towards}, however, it is much more complicated in the case of machine learning models, especially deep neural networks (DNNs), due to factors such as model complexity and the randomness in the training algorithm~\cite{thudi2022unrolling}.

Machine unlearning techniques~\cite{bourtoule2021machine,wu2020deltagrad,guo2020certified,chen2022graph,thudi2022unrolling,gupta2021adaptive,brophy2021machine,thudi2022necessity} have been proposed for erasing training data on machine learning models. 
In machine unlearning, two models are involved in the unlearning process; the original trained model and the unlearned model. 
Existing studies on machine unlearning mainly focus on how to design efficient and effective unlearning methods, that is, how to efficiently obtain the unlearned model while ensuring the information of the unlearned data is removed from the original model. 
However, with two versions of the model available, an adversary has potential resources to infer the private information of the unlearned data. 
In addition, given that machine unlearning is originally proposed to protect the privacy of unlearned samples as a first-class citizen, it is important to understand and investigate the privacy vulnerabilities associated with machine unlearning processes.
These vulnerabilities are underexplored and not captured by the metrics of efficiency and efficacy that govern most of the existing works.

\noindent \textbf{Research Gap.} Within the existing work of machine unlearning, some studies~\cite{marchant2022hard,hu2023duty} start to investigate the vulnerability in machine unlearning. For example, a recent work~\cite{hu2023duty} [NDSS'24] proposes over-unlearning attacks to compromise the utility of the unlearned model. However, from the perspective of privacy, only a few studies~\cite{chen2021machine,wu2020deltagrad,guo2020certified,neel2021descent,carlini2022privacy} mention that machine unlearning may unintentionally compromise the privacy of the unlearned data. 
In the context of DNNs, Chen et al.~\cite{chen2021machine} [CCS'21] specifically demonstrate that the membership privacy of the unlearned samples can be leaked, with membership inference attacks~\cite{shokri2017membership}, through machine unlearning. 
Similarly, Carlini et al.~\cite{carlini2022privacy} observe that a training sample less vulnerable to membership inference attacks could later become vulnerable when other training samples are unlearned. 
However, membership inference attacks focus on inferring the associated information of a data sample, i.e., its membership status on a target model, assuming an adversary already had the sample in hand. 
Thus, membership inference attacks do not capture the full landscape of privacy vulnerability in machine unlearning. 
For example, membership inference attacks cannot reveal the confidential \textit{content} of the unlearned data. 
As such, none of the existing works investigate what confidential or sensitive information contained within the unlearned data can be inferred in machine unlearning.

\noindent \textbf{Research Question.} Given the research gap, we ask the following research question: \textit{``Given access to the original and unlearned models, to what extent do current machine unlearning methods leak the sensitive information of the unlearned data?''}

\noindent \textbf{Motivation.} Machine learning models can be regarded as mappings of their training data~\cite{Goodfellow-et-al-2016,zhang2021understanding}. 
The patterns and relationships present in the training dataset are reflected in the model's parameters and behavior. 
From the perspective of the unlearned samples in machine unlearning, the original model contains the information of these samples because it was trained on them. 
The unlearned model is expected to not contain the information of the unlearned samples because they have been removed during the unlearning process. 
Thus, intuitively, the difference between the original model and the unlearned model reflects the information of the unlearned samples. 
Similarly, this intuition is shared by differencing attacks~\cite{dwork2014algorithmic} in the context of a database.
Here, an adversary can infer the private information of an individual by observing the difference between the answers to two statistical queries on a database.

\begin{figure}[t]
\centering
\includegraphics[width=1.0\linewidth]{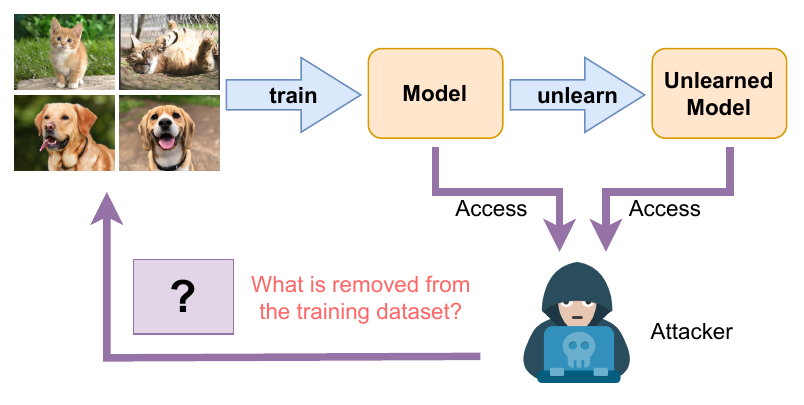}
\caption{An overview of the unlearning inversion attack in machine unlearning. Given access to the original model and the unlearned model, an attacker can mount unlearning inversion attacks to reveal the information of unlearned data.}
\label{fig:attacks}
\end{figure}

To answer the research question, analogous to the model inversion attacks against machine learning~\cite{fredrikson2015model} [CCS'15], we propose \textit{unlearning inversion attacks} to identify the unlearning data leakage in machine unlearning on DNNs, as depicted in \autoref{fig:attacks}.
Specifically, unlearning inversion attacks can reveal two types of private information of an unlearned sample: its feature and its label information when an adversary is given different adversarial knowledge. 
We describe the two types of privacy leakages in more detail.

\noindent \textbf{\textit{Feature Leakage.}} Unlearning inversion attacks can recover the feature of an unlearned sample.
For example, they can recover what an unlearned image looks like. 
To mount this attack, we assume the adversary has white-box access to both the original model and the unlearned model.
This scenario can be satisfied by an honest-but-curious server in Machine Learning as a Service (MLaaS) settings (detailed in Section \ref{sec:threat-model}). 
The intuition behind this attack is that the difference between the two models' parameters provides the gradient information of the unlearned samples. 
Thus, the adversary can adopt gradient inversion techniques \cite{zhu2019deep,zhao2020idlg,geiping2020inverting} to invert the gradient to the corresponding feature of the unlearned sample.

\noindent \textbf{\textit{Label Leakage.}} Unlearning inversion attacks can infer the label of the unlearned data, i.e., inferring which class of samples are unlearned from the original model. In this case, the adversary has only black-box access to the original model and the unlearn model, which can often be satisfied by end users in MLaaS settings (detailed in Section \ref{sec:threat-model}). To mount this attack, the adversary first constructs a few well-crafted probing samples on the original model to capture its behaviour. Then, by querying the unlearned model using the probing samples, the adversary infers the label of the unlearned data based on the difference of the probing samples' predicting outputs on the original model and the unlearned model. The intuition behind the attack is that the prediction output change reflects the model's behaviour change caused by the unlearned data. When unlearning samples of a class, the model will become less confident in predicting the same probing samples into that class. Thus, by observing the prediction difference, the adversary can infer the label of the unlearned data.

\noindent \textbf{Contribution.} Our contributions are summarized as follows:
 
\begin{itemize}[leftmargin=*]

\item 
This paper proposes the first \textit{unlearning inversion attacks} that identify an impact vulnerability in machine unlearning. Through the lens of unlearning inversion attacks, we demonstrate the possibility of obtaining the feature and the label information of the unlearned data in machine unlearning on DNNs. To the best of our knowledge, we are the first to show this possibility, which illuminates a new vulnerability in machine unlearning that has not been explored by existing works.

\item Given access to only the original and unlearned models,
unlearning inversion attacks allow an adversary to obtain the feature and the label information of the unlearned data. When the adversary has white-box access to the models, the features of the removed data can be recovered. 
With only black-box access to the models, the label of the removed data can be inferred. Both two attack scenarios occur in real-world MLaaS environments.

\item We conduct extensive experiments on both exact and approximate machine unlearning methods on benchmark datasets and across various model architectures. 
The findings validate the effectiveness of the proposed unlearning inversion attacks in uncovering the private information in the unlearned data in machine unlearning. 

Three possible defense methods are proposed and evaluated.
However, they lead to unacceptable trade-offs between the defense effectiveness and the utility loss. 

\item The source code and the artifact of the unlearning inversion attacks is released at \url{https://github.com/TASI-LAB/Unlearning-inversion-attacks/tree/main}, which creates a new tool for measuring the privacy vulnerability of machine unlearning methods and sheds light on the design of future unlearning methods.

\end{itemize}

\section{Related Work and Threat Model}
In this section, we first introduce recent studies in machine unlearning domain, including machine unlearning approaches and vulnerability associated within machine unlearning. Then, we describe the threat model of unlearning inversion attacks in MLaaS scenarios. 

\begin{table*}[t]
\centering
\caption{An overview of threats to machine unlearning on DNNs. Our attacks investigate privacy vulnerabilities in machine unlearning, including exact and approximate unlearning, requiring only the knowledge of the model.}
\label{tab:comparison}
\resizebox{\linewidth}{!}{

\begin{tabular}{@{}lcccccl@{}}
\toprule
\multirow{3}{*}{\begin{tabular}[c]{@{}l@{}}\textbf{Attack Type}\end{tabular}} & \multicolumn{3}{c}{\textbf{Adversary's Knowledge Requirement}} & \multicolumn{2}{c}{\textbf{Attack Applicability}} & \multirow{3}{*}{\begin{tabular}[c]{@{}l@{}}\textbf{Vulnerability Type} \end{tabular}}\\ 
\cmidrule(l){2-4} \cmidrule(l){5-6}  
 & \begin{tabular}[c]{@{}c@{}}Knowledge of \\ Dataset\end{tabular} & \begin{tabular}[c]{@{}c@{}}Knowledge of \\ Model\end{tabular} & \begin{tabular}[c]{@{}c@{}}Knowledge of \\ Unlearning Method\end{tabular} & \begin{tabular}[c]{@{}c@{}}Exact \\ unlearning\end{tabular}
 &  \begin{tabular}[c]{@{}c@{}}Approximate \\ unlearning\end{tabular}\\
\midrule
Slow-down unlearning attack~\cite{marchant2022hard} & \pie{360} & \pie{360} & \pie{360} &  \pie{0} & \pie{360} & Efficiency vulnerability\\
Camouflaged poisoning attack~\cite{di2022hidden} & \pie{360} & \pie{360} & \pie{360} & \pie{360} & \pie{0} & Data poisoning vulnerability \\
Over-unlearning attacks~\cite{hu2023duty} & \pie{0} & \pie{360} & \pie{360} & \pie{0} & \pie{360} & Fidelity vulnerability \\ 
Membership inference attacks~\cite{chen2021machine} & \pie{360} & \pie{360} & \pie{360} & \pie{360} & \pie{0}  & Privacy vulnerability \\
\midrule
\begin{tabular}[c]{@{}l@{}}Unlearning inversion attacks (Ours)\end{tabular} & \pie{0} & \pie{360} & \pie{0} & \pie{360} & \pie{360} & Privacy vulnerability\\ 
\bottomrule
\end{tabular}
}
\begin{tablenotes}
\item[] \pie{360}: the knowledge is available or the attack is applicable; \pie{0}: the knowledge is unavailable or the attack is not applicable. 
\end{tablenotes}
\end{table*}

\subsection{Related Work}\label{sec:releated-work}
\noindent \textbf{Machine Unlearning Techniques.} Machine unlearning techniques are motivated by the privacy concerns of individuals and the congruent privacy regulations~\cite{cao2015towards,bourtoule2021machine}. 
A growing number of studies  demonstrate that machine learning models can memorize training data and leak private information contained in the training dataset~\cite{fredrikson2014privacy,fredrikson2015model,carlini2019secret,shokri2017membership}. 
Individuals may feel unhappy with such privacy risks and thus naturally want their data to be deleted from both the training dataset and the machine learning models. 
On the other hand, more and more privacy regulations, e.g., GDPR~\cite{rosen2011right} and CCPA~\cite{pardau2018california}, are issued to protect the privacy of individuals. 
Notably, these regulations grant an individual the ``the right to be forgotten'', which requires companies who have collected and used an individual's data to remove this data  from the trained models once the individual raises deleting requests.

Machine unlearning techniques~\cite{bourtoule2021machine,wu2020deltagrad,guo2020certified,chen2022graph,thudi2022unrolling,gupta2021adaptive,brophy2021machine,thudi2022necessity} are proposed to remove the influence of the removed training data (i.e., the unlearned data) on the trained model.
The techniques can be divided into two groups: exact unlearning and approximate unlearning. 
In exact unlearning, the model is retrained from scratch. 
This involves retraining the model on the training dataset with the unlearned samples removed. 
Exact unlearning has an advantage of completely removing the influence of the unlearned data on the model, but it usually suffers from high computational cost, especially when the model contains high complexity or the training dataset is large. 
To achieve efficient machine unlearning, approximate unlearning techniques are proposed. 
Approximate unlearning is achieved by directly updating the trained model's parameters with information, such as the training gradients, calculated on the unlearned data. 
This typically leads to much lower computational costs compared to exact unlearning. 
However, compared to exact unlearning, approximate unlearning has a weaker unlearning guarantee: the unlearned model may still contain some information of the unlearned data \cite{thudi2022unrolling}.

\noindent \textbf{Machine Unlearning Vulnerability.} 
Despite the progress of machine unlearning techniques towards protecting privacy, recent studies \cite{marchant2022hard,di2022hidden,chen2021machine} show that different kinds of vulnerabilities can exist in machine unlearning.
Marchant et al.~\cite{marchant2022hard} demonstrate an efficiency vulnerability by proposing poisoning attacks that can effectively increase the computation cost of approximate unlearning. 
Specifically, they propose to add well-crafted perturbation to a user's data to create poisoned unlearned samples. 
When unlearning such poisoned samples, complete retraining will be triggered because the error of the approximate unlearning will exceed the allowed error threshold. 
Di et al.~\cite{di2022hidden} demonstrate a data poison vulnerability by proposing targeted attacks that can cause the model to misclassify a particular test sample. 
Specifically, they assume a user can add several well-crafted poisoning samples that contain the information of a target sample in the training dataset of the model before unlearning. 
When the user raises unlearning requests to remove a portion of the well-crafted samples, the model retrained on the remaining training data will misclassify the target sample. 
Hu et al.~\cite{hu2023duty} [NDSS'24] demonstrate a fidelity vulnerability by proposing over-unlearning attacks. 
Specifically, they show that a user can intentionally craft the unlearned data to make the model unlearn more information than expected, leading to a high utility reduction in the unlearned model. 
Chen et al.~\cite{chen2021machine} [CCS'21] demonstrate a privacy vulnerability through membership inference attacks~\cite{shokri2017membership}. 
They show that the membership privacy of the training samples in the original model can be leaked by leveraging the prediction confidence of the unlearned model, even when the membership privacy of such samples is preserved well in the original model. 
Our paper also investigates a privacy vulnerability in machine unlearning. 
However, our work is significantly different from the work~\cite{chen2021machine} in adversary’s knowledge requirements and specific attack goals, which are detailed as follows.

\noindent \textbf{Difference from Existing Attacks.} 
As depicted in \autoref{tab:comparison}, our proposed unlearning inversion attacks differ from existing attacks in machine unlearning in terms of the adversary's knowledge requirements, applicable unlearning type, and the vulnerability type. Compared to Chen et al.~\cite{chen2021machine} [CCS'21] that also investigate privacy vulnerability in machine unlearning, unlearning inversion attacks differ in terms of the adversary's knowledge requirements and specific attack goals. Specifically, the attacker in our attack has knowledge only of the model, while in \cite{chen2021machine} the attacker requires knowledge of the dataset, the model, and the unlearning method. 
In addition, the attacker in \cite{chen2021machine} is assumed to have the unlearned sample at hand, and the attack goal is to launch a membership inference attack to investigate whether such a sample is a member of the original model. 
In our paper, the attacker aims to obtain confidential information contained within the unlearned sample, i.e, its feature or label.
The unlearning inversion attacks are more severe than membership inference attacks. 
We mention that the work~\cite{gao2022deletion} also proposes attacks in machine unlearning to obtain confidential information of the unlearned sample. However, the work~\cite{gao2022deletion} focuses on developing attacks specifically on the K-nearest neighbor model, while we focus on DNNs, which is more widely applicable and has become standard in the field of machine learning.

We note that unlearning inversion attacks share similarities with model inversion attacks \cite{fredrikson2015model,fredrikson2014privacy}, which aim to infer the features of the training data of a machine learning model. 
However, in model inversion attacks, the target model is static (i.e, does not involve two models), and the inferred features are not required to be specific~\cite{dibbo2023sok}.

We note that the work~\cite{salem2020updates} also investigates privacy leakage when two models are involved in a model updating scenario.
The scenario difference between the work~\cite{salem2020updates} and our work is that we focus on model unlearning where existing training data is \textit{removed} from the model, while the work~\cite{salem2020updates} focuses on model updating where newly-collected data is \textit{added} to the model. In addition, our attack method technically diverges from the methodology described in~\cite{salem2020updates} in several key aspects:

\begin{itemize}[leftmargin=*]
    \item \textit{With and without the assumption of a local dataset.} The work~\cite{salem2020updates} assumes that an attacker possesses a local dataset that shares the same distribution as the dataset used to train the target model. This assumption facilitates the attacker in constructing a shadow model that closely mimics the target model's behavior. Following the training of the shadow model, its metadata is harvested to aid in the development of an encoder and a decoder, forming the essential components of the attack mechanism. In contrast, our proposed attack method does not rely on the presumption of access to a local dataset mirroring the training data's distribution (detailed in Section~\ref{sec:attack-method}).
    \item \textit{Feasibility in unlearning cases.} The attack method in~\cite{salem2020updates} relies on prior knowledge of the model's update methodology, including specifics like update rates and updating algorithms. This prerequisite limits its usability to situations where detailed information is readily accessible. In contrast, our attack method operates independently of the underlying unlearning mechanisms, requiring no prior knowledge of specific unlearning methods or parameters. This independence expands the applicability of our approach and enhances its utility in contexts where information about the model's unlearning process may be scarce or inaccessible, underscoring its innovative contribution to the field.
\end{itemize}

\subsection{Our Threat Model}\label{sec:threat-model}
In this subsection, we describe the threat model of unlearning inversion attacks in machine unlearning under MLaaS settings. 
The goal of unlearning inversion attacks is to reveal information in the unlearned data based on access to both the original and unlearned models.

\noindent \textbf{MLaaS Scenario.} Under the settings of MLaaS, there are three entities involved: a model developer, an MLaaS server, and users. The model developer owns a private dataset and is responsible for training the model. Subsequent to the training phase, the developer uploads the trained model to the MLaaS server who deploys the model and provides an API (Application Programming Interface) to users. Through the API, users can send requests to the server and receive responses containing the results of model predictions, i.e., the users can utilize the ability of the model. 

In MLaaS scenarios, machine unlearning happens when some of the samples in the training dataset have to be removed. 
When the unlearning requests to remove such samples have been raised by the individual who contributes the data, the model developer has an obligation to fulfill such requests and remove these samples from the trained model by leveraging a pre-selected unlearning method. 
When the unlearned model is obtained, the model developer uploads it to the server so that the server can replace the existing version of the model on the service platform with the newly unlearned one. 
Then, the users can access the newly deployed model, i.e., the unlearned model, via the API once the replacement is complete. 

While machine unlearning appears to alleviate the individual's privacy concerns by removing her data from the original trained model, we show that machine unlearning can instead leak significant information of the individual to either the server or users in MLaaS scenarios. 
Specifically, we propose unlearning inversion attacks that enable the server and a user to recover the feature and the label of the unlearned data. 
Before we introduce the details of the unlearning inversion attacks, we detail the capabilities and knowledge of the two attackers.

\noindent \textbf{Server as an Attacker.} 
We assume the server is honest-but-curious: the server honestly hosts and provides machine learning services to users but may still have the potential to be curious or inadvertently gain the knowledge about the unlearned data. 
To simulate a practical scenario, we assume the server has only the model parameters uploaded from the model developer, i.e., white-box access to the original and unlearned models, but without any additional prior knowledge such as either the knowledge of the unlearning methods used by the model developer or the training data distribution. 

\noindent \textbf{User as an Attacker.} 
The users interact with the server's machine learning model through the provided API. 
Consequently, the user has only black-box access to both the original and unlearned models, i.e., the user submits a sample to the server and receives the prediction output of the model. 
Note that the black-box assumption is the most difficult setting for an attacker. 

We note that the server might be compromised by the user so that the user can infer more information of the unlearned sample. 
In this paper, we do not consider this scenario. 
This is because we focus on investigating how machine unlearning in MLaaS settings naturally leaks the information of the unlearned sample. 
Specifically, we assume that the adversarial knowledge of the user-attacker is identical to the knowledge of a normal user. 
Investigating how a malicious user can infer more than the label information of the unlearned sample, by assuming the server can be compromised, is orthogonal to our work.

\section{Methodology}

\begin{figure*}[t]
\centering
\includegraphics[width=0.95\linewidth]{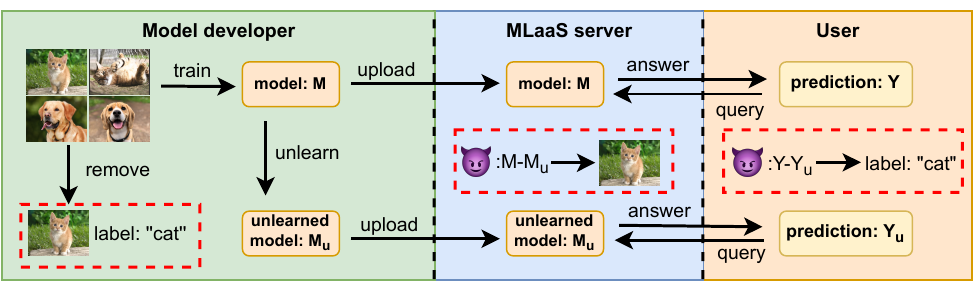}
\caption{An overview of unlearning inversion attacks in the MLaaS environment. The server leverages the differences between two model parameters to recover the unlearned sample's feature. The user uses the differences between two models' prediction outputs to infer the unlearned sample's label.}
\label{fig:unlearn-leak}
\end{figure*}
In the context of machine learning, a model reflects its training data through the learning process, during which the model adjusts its parameters to capture patterns and relationships present in the training data \cite{goodfellow2016deep}. 
In other words, the model parameters and their resulting behavior contain information about the training data. 
In machine unlearning, the original model contains information about the removed data, as the model was trained on it.
In contrast, the unlearned model does not contain this information (there might be some residual information contained in the model depending on the specific unlearning method \cite{guo2020certified}). 
Note that, omitting the randomness in the training algorithm, the unlearned model differs from the original model only in the unlearned data. 
Thus, intuitively, the difference between the original model and the unlearned model encodes the specific information of the unlearned data. 
The challenge is then how to decode this difference between the two models into specific patterns contained in the unlearned data.

In this section, we introduce two methods that an attacker can leverage for mounting unlearning inversion attacks in MLaaS scenarios.
The methods constitute a response to the research question raised in Section~\ref{sec:introduction}.
The first method enables the server to recover the feature of the unlearned data, while the second method allows the user to infer the label of the unlearned data.

\subsection{Problem Statement}
In this paper, we mainly focus on machine learning classification because it is one of the most widely used applications.
Existing studies of machine unlearning mainly focus on this task. 
We define a data sample as $(\bm{x},y)$, where $\bm{x}$ is its multidimensional feature and $y$ is its class label. 
A machine learning classifier is a function $f(\cdot)$ that maps input data to output predictions, i.e., it takes as input $\bm{x}$ and outputs $\mathcal{Y}=f(\bm{x})$, which is a vector of probabilities with length equal to the number of classes in the classification task. 
In $\mathcal{Y}$, each entry $y_i$ indicates the posterior probability of the classifier assigning $\bm{x}$ to a class $c_i \in \mathcal{C}$, where $\mathcal{C}$ is the set of all classes.
We let $\mathcal{D}$ be a training dataset that contains the samples used for training the machine learning model.
Let $\mathcal{D}_{u} \subseteq \mathcal{D}$ be a dataset that contains the unlearned samples that need to be removed. 
We define $\bm{\theta}$ as the original model parameter trained on $\mathcal{D}$ and $\bm{\theta}_u$ as the unlearned model parameter that removed the influence of $\mathcal{D}_u$ from $\bm{\theta}$ through an unlearning method. 
The goal of unlearning inversion attacks is to reveal the information in the unlearned sample based on access to both $\bm{\theta}$ and $\bm{\theta}_u$.

\subsection{Unlearning Inversion Attacks}\label{sec:attack-method}
\autoref{fig:unlearn-leak} shows an overview of unlearning inversion attacks in the MLaaS scenario. 
At the beginning, a model developer trains a model on the training dataset and uploads the trained model $M$ to an MLaaS server for deployment. 
After the model is deployed, users can query the model and receive corresponding predictions. 
When a sample in the training dataset needs to be removed, the developer deletes it from the training dataset in addition to generating an unlearned model $M_u$, which is uploaded to the server for replacing the original model $M$. 
After the deployment of the unlearned model is finished, the user can utilize the new model for prediction. 
Unlearning inversion attacks occur when $M_u$ is uploaded to the server and is accessible to the user. 
Specifically, the server leverages the difference between the two models' parameters for feature recovery, while the user utilize the difference between the prediction outputs of the two models for label inference. 
We now detail both types of information leakage.

\subsubsection{Unlearning Inversion Attacks for Feature Recovery} 
The server mounts the unlearning inversion attack to recover a sample $\bm{x}$ that is close to the unlearned sample $\bm{x}^* \in \mathcal{D}_u$. 
To perform this attack, there are two steps: \textit{i)} obtain the gradient information of the unlearned sample by leveraging the difference between the parameters of the original and unlearned models; \textit{ii)} decode the gradient information to $\bm{x}$ through an optimization algorithm. 
We detail the two steps as follows.

\noindent \textbf{Gradient Estimation.} 
From the perspective of a data sample, its contribution to model training is reflected in the gradient of its loss with respect to the model parameters. 
During the training process, each data point in the training dataset contributes to the overall update of the model's parameters. 
Model parameter updates usually adhere to the following steps:
\begin{equation}
    \bm{\theta}_{k+1}=\bm{\theta}_{k}-\alpha \nabla_{\bm{\theta}} \mathcal{L}(\bm{\theta}_k),
\end{equation}
where $\bm{\theta}_{k}$ represents the model parameters at iteration $k$, $\bm{\theta}_{k+1}$ represents the new parameters after the update, $\alpha$ is the learning rate, and $\nabla_{\bm{\theta}} \mathcal{L}(\bm{\theta}_k)$ is the gradient of the loss function $\mathcal{L}(\cdot)$, calculated on a batch of training examples $\mathcal{D}_{b}$, with respect to the model parameters $\bm{\theta}_k$. 
Intuitively, the difference between two model parameters, i.e., $\bm{\theta}_{k+1}$ and $\bm{\theta}_k$, represents an approximation of the gradient information. 
Notably, because $\bm{\theta}_{k+1}$ differs from $\bm{\theta}_k$ in an update using the gradient of $\mathcal{D}_{b}$, the difference between $\bm{\theta}_{k+1}$ and $\bm{\theta}_k$ is an estimation of the gradient information of $\mathcal{D}_{b}$.

In machine unlearning, the original model $\bm{\theta}$ contains the information of $\mathcal{D}_u$, while the unlearned model $\bm{\theta}_u$ does not contain this information.
In addition, omitting the randomness associated with the training algorithm, $\bm{\theta}_u$ only differs from $\bm{\theta}$ in the unlearning samples that had been removed from $\bm{\theta}$.
Thus, the difference between $\bm{\theta}$ and $\bm{\theta}_u$ gives an estimation of the gradient information of $\mathcal{D}_u$, which is formulated as:
\begin{equation}
    \nabla^* = \bm{\theta}-\bm{\theta}_u.
\end{equation}

The challenge of the attack now becomes how to decode such gradient information to the unlearned data.

\noindent \textbf{Gradient Inversion.}  In the context of federated learning \cite{mcmahan2017communication}, many recent studies \cite{zhu2019deep,zhao2020idlg,geiping2020inverting,zhu2020r,yin2021see,yue2023gradient} have shown that federated learning is vulnerable to gradient inversion attacks. 
Specifically, these studies demonstrate that sharing the gradient with the parameter server in federated learning cannot protect the privacy of local clients because the parameter server can leverage optimization algorithms to reconstruct the private training data from the gradient. 
Motivated by such gradient inversion attacks, we propose to leverage optimization algorithms to decode the estimated gradient of the unlearned data to the unlearned data. 

However, not all optimization algorithms developed in gradient inversion attacks can be adopted to our unlearning inversion attacks. 
First, in gradient inversion attacks, the gradient used as the optimization target is exactly the gradient of the training data with respect to the model parameters. 
However, as we assume the adversary does not know the details of the unlearning method in our unlearning scenario, we can only obtain an estimation of the gradient of the unlearned data rather than its exact value. 
Second, some gradient inversion attacks require more knowledge than the gradient information of the data to reconstruct the data. 
For example, \cite{yue2023gradient} requires knowledge of the training data distribution for creating a synthetic dataset, which is used to train a neural network for postprocessing the reconstructed data. 
In our scenario, we assume the attacker has only model parameters to estimate the gradient without additional prior knowledge such as the training data distribution. 

To overcome the challenge that the adversary has only an estimation of the gradient, we adopt the optimization algorithm in \cite{geiping2020inverting} to recover the unlearned data from the estimated gradient. Specifically, we let $\bm{x}'$ be a dummy input and $y'$ be a dummy label. Based on the original model parameter $\bm{\theta}$, we obtain the gradient of the dummy sample:
\begin{equation}
    \nabla'=\frac{\partial \mathcal{L}\left(f_{\bm{\theta}}(\bm{x}'), {y}^{\prime}\right)}{\partial \bm{\theta}}.
\end{equation}
Then, the cost function defined for optimizing the dummy variables $\bm{x}'$ and $y'$ is:
\begin{equation}\label{equ:cost-function}
\arg \min _{\bm{x}'} -l(\nabla',\nabla^*)+\alpha \textrm{TV}(\bm{x}'),
\end{equation}
where $l(\nabla',\nabla^*)$ is the cosine similarity defined as:
\begin{equation}
l(\nabla',\nabla^*)=\frac{\left\langle \nabla', \nabla^* \right\rangle}{\left\|\nabla'\right\|_2 \left\|\nabla^*\right\|_2}.
\end{equation}
$\textrm{TV}(\bm{x}')$ represents the total variation of $\bm{x}'$ considering $\bm{x}'$ as an image:
\begin{equation}
    \textrm{TV}(\bm{x'})=\sum_{i, j \in {N}}\left\|{c}_i-{c}_j\right\|_1.
\end{equation}
Here $N$ defines the pixel neighborhood of the image, $c_i$ denotes the pixel value, $\left\|\cdot\right\|_1$ is the 1-norm, TV is used as a basic image prior~\cite{wang2019beyond} and $\alpha$ is a hyperparameter balancing two terms. 

As $\nabla^*$ is not the exact gradient values of the unlearned data, the advantage of the cost function in Equation \ref{equ:cost-function} is that it aims to obtain a dummy gradient $\nabla'$ that matches the \textit{direction} of the target gradient $\nabla^*$ rather than matching its magnitude.
Accordingly, we can obtain a dummy sample $\bm{x'}$ that has a gradient direction close to that of the unlearned sample $\bm{x^*}$. 
In the experiments, we will demonstrate that under this cost function, $\bm{x}'$ and $\bm{x}^*$ are also close in the pixel space.

In the optimization process, we assume the label information of the unlearned sample is known and focus on recovering the feature as it carries significant private information. 
We make this assumption because existing studies~\cite{yin2021see,wainakh2021label} have designed efficient algorithms to extract the label of the data based on the sign of the gradient, which is known to the attacker in our scenario. 
In addition, this assumption is based on the attacker's ability to
systematically test each possible label until a meaningful feature is recovered. Despite higher computational costs, our attacks can still operate with
unknown labels by treating them as variables. 
The recovery results in Figure~\ref{fig:label-unknown} in Appendix~\ref{subsec:additional_experiments} reveal comparable outcomes for both known and unknown label cases.

\subsubsection{Unlearning Inversion Attacks for Label Inference} Under the MLaaS setting, the user has only black-box access to the original and unlearned models.
This makes it difficult for the user to leverage the model parameters to recover the features of the unlearned data.
However, in this case, we demonstrate that the user can still perform unlearning inversion attacks to reveal the information in the unlearned data by leveraging the prediction output of the models. 
Specifically, the user is able to infer the label of the unlearned data by leveraging the difference between the perdition output of the original model and that of the unlearned model. 
To perform the attack, two steps are required: \textit{i)} construct probing samples to capture the prediction behaviour of the original model; \textit{ii)} probe the unlearned model to infer the label of the unlearned data.

\noindent \textbf{Constructing Probing Samples.} In classification tasks, the model's prediction output is a key reflection of its behavior, indicating the model's assignment 
of class labels based on the learned patterns in the training data. For the same sample, the prediction output of the sample produced by the unlearned model is highly likely to be different from that of the original model, as patterns in the unlearned model's training data have changed. The challenge is how to decode such a prediction change to the label information of the unlearned data. To solve this challenge, we consider the user can construct a few well-crafted samples (denoted as probing samples) on the original model to capture its prediction behaviour. Then, the user can leverage these probing samples to probe the unlearned model for inferring the label information.

The attacker creates a dataset $\mathcal{D}_p$ that contains $n$ samples $\bm{x}_1,\cdots,\bm{x}_n$. 
As the attacker does not know the training data distribution, these samples can be from a different data distribution than that of the training data of the model.
Let $C$ denote the number of classes in the classification task.
Since the attacker has black-box access to the model, for each $\bm{x} \in \mathcal{D}_p$, the attacker can obtain the prediction output $\mathcal{Y}= [p_1,\cdots,p_C]$. 
Given a sample $\bm{x} \in \mathcal{D}_p$, the attacker aims to add perturbation $\bm{\delta}$ to it so that the original model $\bm{\theta}$ assigns $\bm{x}'=\bm{x}+\bm{\delta}$ to a particular class $y_t$ with high confidence. 
We define the loss function for optimizing the sample $\bm{x}' \in \mathbb{R}^{N}$ (considering $\bm{x}'$ is an image) as follows:
\begin{equation}
\begin{aligned}
& \arg \min _{\bm{x}'} g(\bm{x}',y_t) \\
& \text {subject to } \bm{x}' \in[0,1]^N,
\end{aligned}
\end{equation}
where 
\begin{equation}
    g(\bm{x}', y_t)= \max _{i \neq y_t}[Z(\bm{x}')]_i-[Z(\bm{x}')]_{y_t},
\end{equation}
where $Z(\bm{x}')$ are the logits of the original model and $[Z(\bm{x}')]_i$ is the predicted probability that the original model assigns $\bm{x'}$ to the class $i$.

The goal of making the model predict a sample towards a target label is similar to adversarial attacks \cite{carlini2017towards,papernot2016limitations,papernot2017practical}. 
Given that the attacker in our setting has only black-box access to the original model, we utilize the black-box adversarial example technique of the zeroth order optimization (ZOO) based attack \cite{chen2017zoo}. 
We select this technique because it not only helps us create a sample $\bm{x}'$ with a target predicted label, but also enables us to obtain $\bm{x}'$ towards the target predicted label with high confidence. 
We mention that other black-box adversarial example techniques such as HopSkipJump \cite{chen2020hopskipjumpattack} and QEBA \cite{li2020qeba} may also be considered for constructing the probing samples in this step.

To comprehensively capture the original model's behaviour, for each label $y_t \in \mathcal{C}$, we construct $m$ probing samples and record their corresponding prediction vector $\hat{\mathcal{Y}}=f_{\bm{\theta}}(\bm{x}')$. Note that the number of the probing samples should be small to save the effort of the attacker. In the experiments, we demonstrate that a small number of probing samples, e.g., 10 for each class, is enough for label inference. We call the dataset $\mathcal{D}_p$ as the probing dataset when such probing samples are constructed. The probing dataset helps to capture the changes in the unlearned model’s training data based on its prediction output.

\noindent \textbf{Probing the Unlearned Model.} After constructing the probing dataset, the attacker uses it to probe the unlearned model when it is deployed. For each $\bm{x}'_i \in \mathcal{D}_p$ that the original model predicts as class $y_t$, we record $\mathcal{Y}_i=f_{\bm{\theta}_u}(\bm{x}')$ predicted by the unlearned model. We calculate the average difference of the outputs of the original model and the unlearned model:
\begin{equation}
    \Delta_p=\frac{1}{m}\sum_{i=1}^m (\hat{\mathcal{Y}_i}-\mathcal{Y}_i),
\end{equation}
\begin{equation}
    \beta_{y_t}=\Delta_p[y_t],
\end{equation}
where $\hat{\mathcal{Y}_i}$ is the prediction output of the original model on the probing sample $\bm{x}'_i$.

Intuitively, $\beta_{y_t}$ captures the prediction change on the class $y_t$ that the unlearned model behaves on the probing samples. 
Note that the original model assigns a high prediction confidence (close to 1) to a target class for the probing samples. 
If a certain number of training samples of a class is removed from the model, the model should be less confident in assigning that class label for the same probing samples, because the model has less information to rely on when making predictions for samples of that class. 
Based on this intuition, the attacker can check which class in the probing samples has the largest confidence drop to infer the label of the unlearning data.

Formally, to determine which class the unlearned data belong to, the attacker finds the class index that has the largest prediction confidence drop:
\begin{equation}\label{equ:label}
    \mathcal{C}_u=\arg \max _i \beta_i.
\end{equation}

Note that Equation \ref{equ:label} works for the case where samples of a single class are removed from the model. 
In complicated unlearning scenarios, the original model can unlearn samples from different classes (assuming $K$ different classes) at the same time. 
Our strategies can also be applied in this case: the attacker finds the $K$ class indexes that have the top-$K$ largest confidence drop in the probing samples, and infer these $K$ classes as the labels of the unlearned data.

\section{Experimental Settings}
In this section, we first introduce the datasets, models, and evaluation metrics used in the experiments. Then, we introduce the unlearning settings, including the unlearning methods in evaluating unlearning inversion attacks.

\subsection{Datasets and Models}
\noindent \textbf{Datasets.} We use three benchmark datasets of CIFAR-10~\cite{krizhevsky2009learning}, CIFAR-100~\cite{krizhevsky2009learning}, and STL-10~\cite{coates2011analysis} to evaluate the proposed unlearning inversion attacks in the experiments. The three datasets are widely used for image classification tasks and are widely used in existing machine unlearning studies.
They cover a wide range of object categories with different learning complexities.
In addition, to fully capture the diversity of challenges in machine unlearning across different domains, we also conducted experiments on two datasets: Chest X-Rays~\footnote{\url{https://www.kaggle.com/datasets/paultimothymooney/chest-xray-pneumonia/data}} from the medical domain and the German Credit dataset~\footnote{\url{https://archive.ics.uci.edu/dataset/144/statlog+german+credit+data}} from the finance domain. These two datasets help to study the impact of unlearning inversion attacks on models trained on highly sensitive or dimensional data. The experimental results on Chest X-Ray and Germany Credit are provided in Appendix~\ref{subsec:additional_experiments}. A detailed description of the five datasets is provided in the Appendix \ref{appendix:settigns}.

For each dataset, we spilt the training dataset into two disjoint dataset with 80\% of the training samples in $\mathcal{D}_0$ and 20\% training samples in $\mathcal{D}_u$. We simulate $\mathcal{D}_0$ as a public dataset used for training a pre-trained model and $\mathcal{D}_u$ as a private dataset that contains the training data owned by the model developer. In our experiments, we first train the model on $\mathcal{D}_0$ to obtain $M_0$, which acts as the pre-trained model. Then, we fine-tune $M_0$ on $\mathcal{D}_u$ for several epochs to obtain the final trained model $M$. We aim to simulate the scenario where the model developer employs a pre-trained model trained on a public dataset as a feature extractor, which is popular in model training of DNNs and usually leads to better model performance \cite{carreira2017quo,zhao2017pyramid,he2019rethinking}. This setting also aligns with existing machine unlearning literature \cite{guo2020certified,thudi2022unrolling,wu2020deltagrad} and has been demonstrated to help to reduce the unlearning error on the unlearned model \cite{thudi2022unrolling}.

\noindent \textbf{Models.}
We select three model architectures of different sizes in our experiments.
Specifically, we use a 7-layer convolutional neural network, denoted as ``ConvNet'' for CIFAR-10 and CIFAR-100, and ResNet-18 for STL-10 because of its larger image size.
In addition, we adapt ConvNet to single-channel dataset Chest-Xray and use a 5-layer multi-layer perceptron (MLP) connected by ReLU for Germany Credit.
To ensure the quality of the pre-trained model, we train the model on the public dataset $\mathcal{D}_0$ for 120 epochs using the SGD optimizer of weight decay 5e-4 and a initial learning rate 0.1 linearly decayed with epochs. After obtaining the pre-trained model, we fine-tune the pre-trained model on the private training dataset $\mathcal{D}_u$ with a batch size of 128 and a learning rate 0.001 for several epochs.

\noindent \textbf{Metrics.} In unlearning inversion attacks for feature recovery, we use the metrics of MSE, PSNR, and LPIPS to measure the recovery quality following existing works of reconstruction attacks~\cite{geiping2020inverting, yue2023gradient}. In general, smaller values of the MSE and LPIPS, or higher PSNR values represent better quality of the recovered features. The detailed description of the three metrics is introduced in Appendix \ref{appendix:settigns}. Note that the three metrics only serve as a reference for measuring the quality of the recovered feature in unlearning inversion attacks, i.e., they measure how close a recovered feature is to the original feature. In some cases of the following experiments (See Section \ref{subsec:abla_feature_recoverty}), we will show that there exist recovered examples of high metric distance. However, such examples can still leak identifiable object similar to that in the unlearned sample. In unlearning inversion attacks for label inference, we use the accuracy to measure the attack performance. Specifically, we examine whether the inferred label matches the actual label of the unlearned data.

\begin{figure*}[t]
    \centering
    \includegraphics[width=0.9\linewidth]{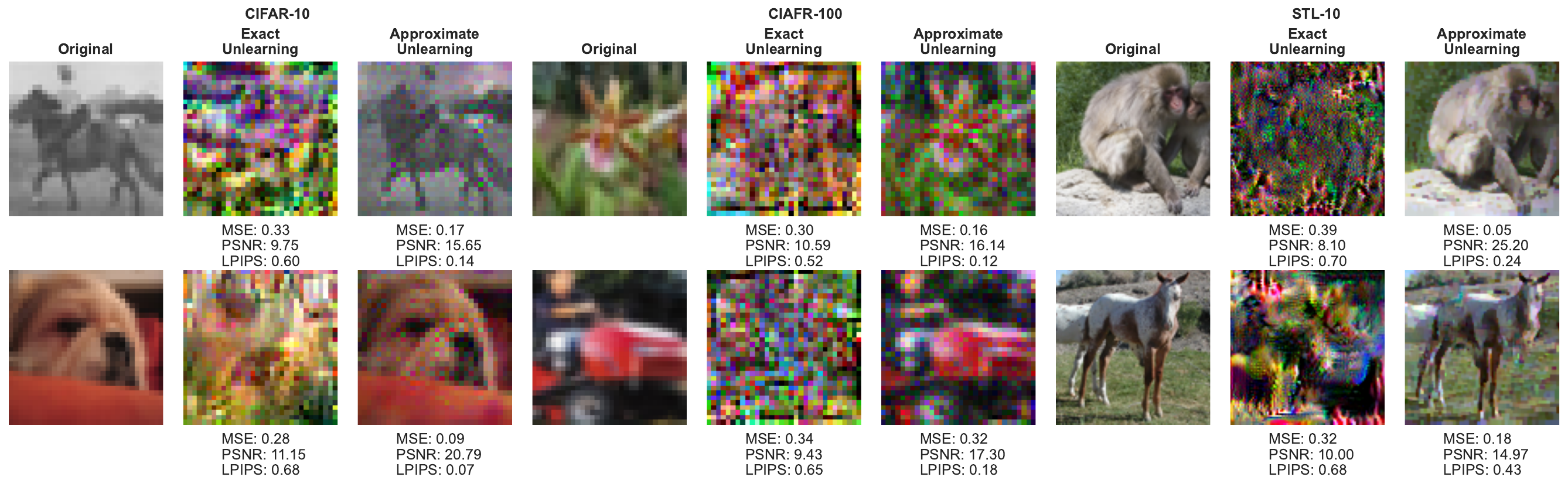}
    \caption{The recovered data from the exact and approximate unlearning on CIFAR-10, CIFAR-100, and STL-10.}
    \label{fig:recons_examples}
\end{figure*}

\subsection{Machine Unlearning Settings}
\noindent \textbf{Machine Unlearning Method.} To conduct a comprehensive investigation on the privacy vulnerability in machine unlearning, we consider unlearning inversion attacks on both exact unlearning and approximate unlearning methods. As there exist many unlearning methods, we select two representative unlearning methods for evaluation and report the attack performance. In this paper, our experimental results serve as an initiation of the exploration of the privacy vulnerability in machine unlearning. We note that the privacy vulnerability investigation on other unlearning methods through the lens of unlearning inversion attacks is orthogonal to our work. In our experiments, we select retraining as the exact unlearning method and the single gradient unlearning method \cite{thudi2022unrolling} as the approximate unlearning method. We detailed introduce the two unlearning approaches and describe the reasons for choosing them in Appendix \ref{appendix:settigns}.

\noindent \textbf{Unlearning Setting.} In machine unlearning, the unlearned data may consist of multiple samples, and they can be from different classes. To simplify the experiments, in unlearning inversion attacks for feature recovery, we start from the case where the original model and the unlearned model only differ from one sample. In unlearning inversion attacks for label inference, we start from the case where samples of one class in the unlearned data are removed. The two cases aim to show the information of the unlearned data can indeed be leaked in machine unlearning, as depicted in Section \ref{sec:effectiveness}. In Section \ref{sec:ablation-study}, we conduct a comprehensive investigation on how different factors in machine unlearning, such as the number of unlearned samples and classes, affect the performance of unlearning inversion attacks.

\section{Evaluation}\label{sec:effectiveness}

\subsection{Effectiveness of Feature Recovery}\label{subsec:effectiveness_of_feature_recovery}

To begin with, we present the feature recovery results when unlearning a single sample from the original model while providing the case of unlearning multiple samples in Section \ref{sec:ablation-study}.
In Figure~\ref{fig:recons_examples}, we show the original samples and the recovered samples in exact unlearning and approximate unlearning on CIFAR-10, CIFAR-100, and STL-10 when setting the fine-tune epoch as one.
Along with the visualization, we quantify the visual distance between the recovered and the original samples using previously described metrics.

We can see that the unlearned samples can be correctly recovered with sufficient details to identify the objects in approximate unlearning.
Although the unlearned data cannot be clearly recovered from exact unlearning, the unlearned model in exact unlearning needs to be retrained, which can be impractically expensive.
On the other hand, approximate unlearning as a more efficient alternative, the unlearned data can be clearly recovered.  
Even though the MSE and LPIPS can be high for some recovered examples (e.g., the leftmost example of the second row (i.e., the ``dog'' image) in Figure~\ref{fig:recons_examples} in approximate unlearning, the main object is still recognizable from the recovered feature. 

In general, feature recovery for exact unlearning is often more challenging than for approximate unlearning, despite the potential for a larger model difference in the former. 
This is due to the presence of extraneous randomness in retraining, leading to inaccuracies in mapping the difference to the feature space. However, exact
unlearning may raise privacy concerns, especially in scenarios with minimal training data, such as few-shot learning. In Section~\ref{subsec:abla_feature_recoverty}, we will show that when the training dataset is small, feature recovery in exact unlearning can also be successful, and the attack performance is close to that in approximate unlearning.  Additional experiments of successful feature recovery in exact unlearning are provided in Figure~\ref{fig:effective-single-unlearn} in the Appendix~\ref{subsec:additional_experiments}.

\subsection{Effectiveness of Label Inference}
To infer the label of the unlearned data, we construct 10 probing samples per class on the original model.
We use random Gaussian noise as an initialization of the probing sample, and leverage the ZOO technique to maximize the prediction confidence of the probing samples for their targeted class.
We optimize the sample until its prediction confidence is higher than an early-stopping threshold.
We set the stopping threshold to $1-1 \times 10^{-3}$ so that the probing samples are predicted to a target class with confidence close to 1 on the original model.
In this experiment, we mainly consider STL-10 because it has higher-resolution images compared to CIFAR datasets, providing more detailed visual information. STL-10 helps simulate common MLaaS API services that provide detailed image analysis. To obtain the original model, we fine-tune the pre-trained model on a sub-training dataset $\mathcal{X}_u \in \mathcal{D}_u$. The subset contains 512 samples with around 50 samples per class, which helps to understand how a small number of the unlearned samples in a class can affect the attack performance. 

\begin{figure}[t]
    \centering
    \includegraphics[width=0.75\linewidth]{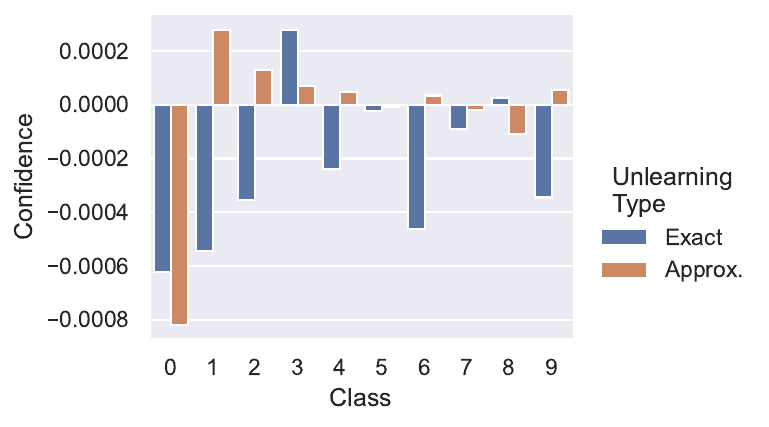}
    \caption{The prediction confidence changes of probing samples for exact unlearning and approximate unlearning. The class with a bar below 0 with the largest height represents the inferred label of the unlearned data.}
    \label{fig:label_infer_oneclass}
\end{figure}

In Figure~\ref{fig:label_infer_oneclass}, we plot the confidence changes of the probing samples between the unlearned model and the original model.
Without loss of generality, we consider the original model unlearns data samples of class 0 and show results of unlearning partial class-level data and data of other classes in Section~\ref{sec:ablation-study}.
The blue and orange bars represent the confidence change in each class in exact unlearning and approximate unlearning, respectively.

\noindent\textbf{Effectiveness for Approximate Unlearning.}
From Figure~\ref{fig:label_infer_oneclass}, we can see that the orange bar has the lowest value on class 0, showing that approximate unlearning causes maximum confidence drop of the probing samples on class 0. This enables the adversary to correctly infer the label of the unlearned data as class 0.
Another two confidence drop is in class 7 and class 8, but the magnitude is far lower than that of class 0. Thus, the adversary can proceed with the label inference in this case with high certainty.

\noindent\textbf{Effectiveness for Exact Unlearning.}
In Figure~\ref{fig:label_infer_oneclass}, the blue bar has the lowest negative value on class 0, indicating the adversary can also correctly infer the actual label of class 0 of the unlearned data in exact unlearning.
However, compared to approximate unlearning, the confidence drops for other classes, such as class 1 and class 6, are closer to that of class 0.
We suspect the reason is that the close representation of samples between certain classes makes the original model's decision boundary shift similarly for such correlated classes during the unlearning process.
For example, class 0 is ``airplane,'' and the class with the second-highest confidence drop is class 1 of ``birds,'' which can be similar to the ``airplane'' (e.g., images of sky background).
Accordingly, when class 0 is removed, the model becomes less confident on samples of class 1, as samples of class 0 may assist the model's learning of class 1.

\begin{mdframed}[backgroundcolor=white!10,rightline=true,leftline=true,topline=true,bottomline=true,roundcorner=2mm,everyline=true]
\textbf{Takeaway 1.~}
Unlearning inversion attacks are effective in machine unlearning. For approximate unlearning, the attack can recover the feature as well as the label information of the unlearned data. For exact unlearning, the attack is effective for label inference but is less effective in feature recovery.
\end{mdframed}

\section{Ablation Study}\label{sec:ablation-study}
In this section, we conduct a comprehensive ablation study to investigate how different factors in machine unlearning affect the performance of our proposed attack.

\subsection{Ablation Study for Feature Recovery}\label{subsec:abla_feature_recoverty}
We first investigate how different factors can affect unlearning inversion attacks for feature recovery.

\noindent\textbf{Training Dataset Size.}
In Section \ref{sec:effectiveness}, we have demonstrated the effectiveness of unlearning inversion attacks for feature recovery in approximate unlearning while not observing that in exact unlearning. Here, we show that when the training dataset size is small, unlearning inversion attacks for feature recovery can also be successful in exact unlearning. To conduct the experiments, we consider a small training dataset with a size of 4, 8, and 16 sampled from the images in $\mathcal{D}_u$ of CIFAR-10. We repeat 10 times the sampling process using different random seeds for each data size. For machine unlearning, we consider the original model and the unlearned model differ only from one sample. Although the dataset size is quite small here and may not be practical, our experimental results serve as an exploration of the extreme cases of privacy vulnerability in exact unlearning, which can call attention to MLaaS applications where training samples are very limited.

\begin{figure}
    \centering
    \includegraphics[width=0.9\linewidth]{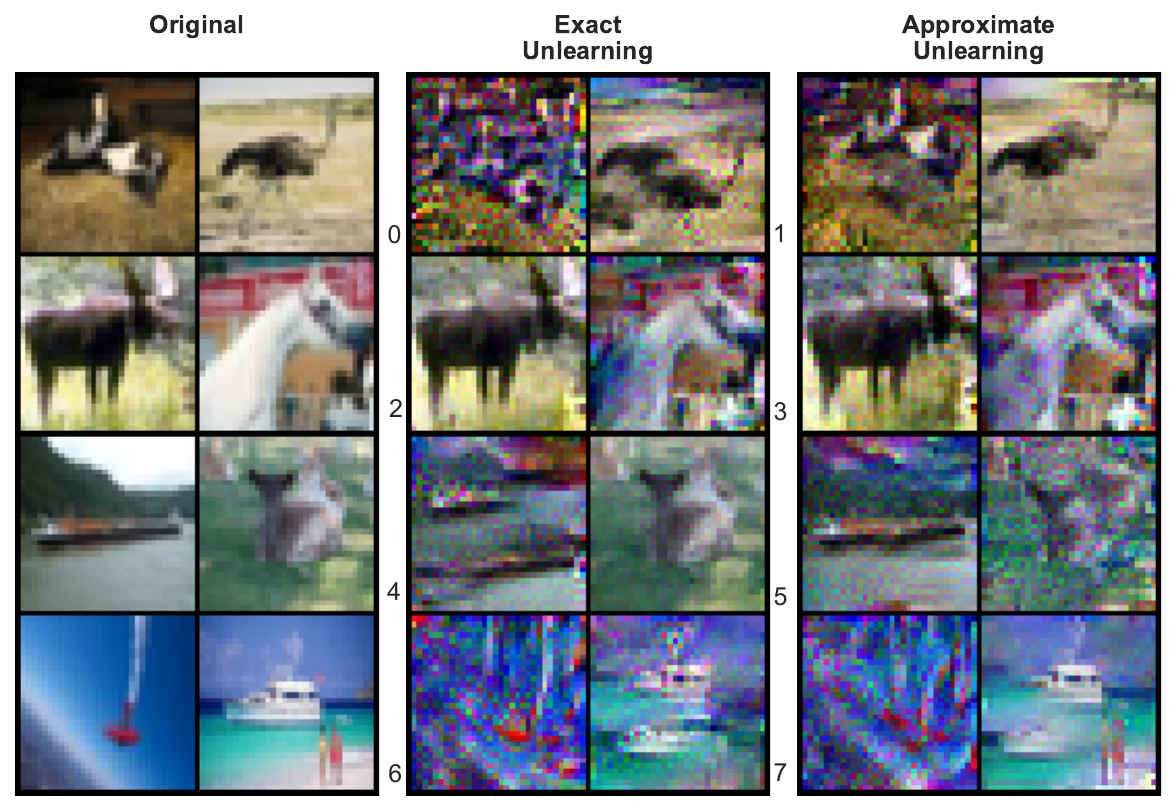}
    \caption{Feature recovery when the training dataset size is 8. The image IDs along the middle column are used to mark specific image positions for the presentation of Table~\ref{tab:metric_stat}.}
    \label{fig:cifar10_batch8}
\end{figure}

In Figure~\ref{fig:cifar10_batch8}, we visualize the recovered features in both exact unlearning and approximate unlearning with a dataset size of 8. As we can see, feature recovery is effective in approximate unlearning when the training dataset is small. Together with the findings in Section~\ref{sec:effectiveness}, this demonstrates that privacy vulnerability is severe in approximate unlearning. For exact unlearning, some of the objects in the recovered images become recognizable and are similar to the objects in the features of the unlearned images, e.g., the ``deer'' in image ID 2. This complements the findings in Section~\ref{sec:effectiveness} and demonstrates that exact unlearning may also leak feature information of the unlearned data in the extreme case of a small training dataset. 

\begin{figure}[t]
    \centering
    \includegraphics[width=0.9\linewidth]{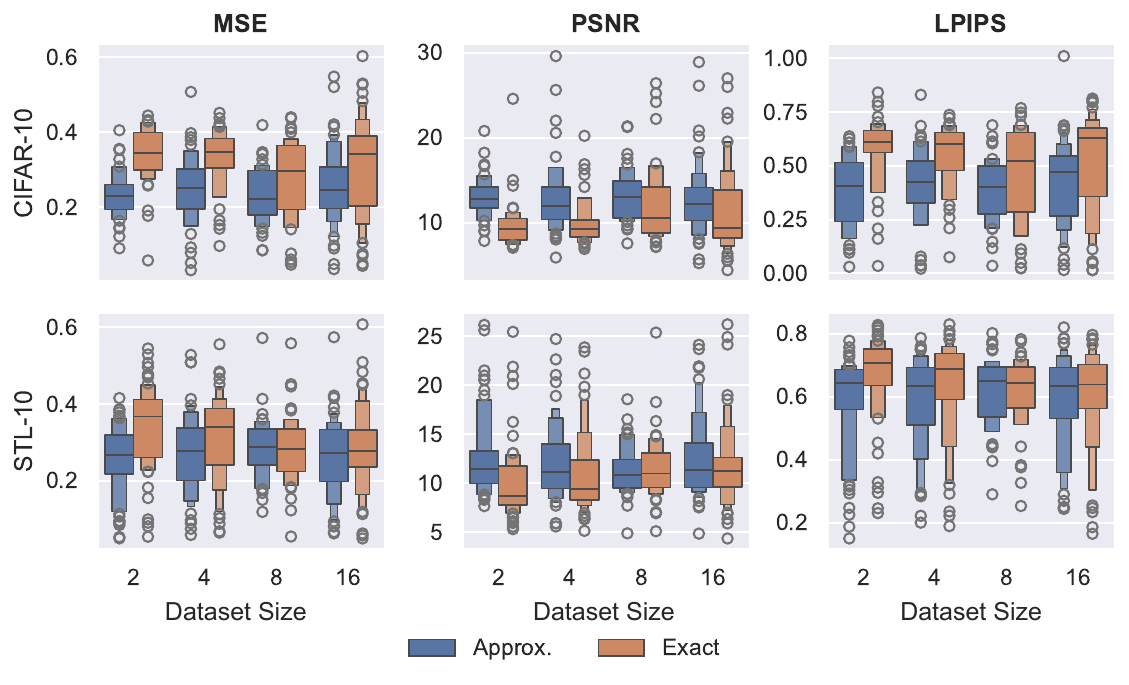}
    \caption{The metrics of recovered features when setting different dataset sizes.}
    \label{fig:reconst_quality_onesample}
\end{figure}

Now, we are aware that with a small training dataset, unlearning inversion attacks for feature recovery are effective in both exact unlearning and approximate unlearning. Figure~\ref{fig:reconst_quality_onesample} shows how the size of the dataset affects the quality of the recovered feature. As depicted, the quality of the recovered feature does not worsen as the dataset size increases because the distances between the unlearned data and the recovered data are similarly distributed.

We also observe that some unlearned data can be clearly recovered, as indicated by the round points of low perception distance in Figure~\ref{fig:reconst_quality_onesample}. This phenomenon suggests that some unlearning samples are more vulnerable to unlearning inversion attacks, which is further discussed in Section \ref{sec:discussion}.

One interesting finding is that approximate unlearning has generally better recovery results compared to exact unlearning, which demonstrates that approximate unlearning leaks more information than exact unlearning.
This is expected because approximate unlearning directly modifies the original model: the original model is updated with one step to obtain the unlearned model. Thus, the weight difference between the two models can well capture the gradient information of the unlearned data. By contrast, in exact unlearning, the unlearned model is retrained, during which many randomnesses can exist, e.g., the randomness in the training algorithm, leading to a biased estimation of the gradient.

Note that the metrics of MSE, PNSR, and LPIPS only serve as a reference for measuring the recovery quality, i.e., measuring how close the recovered feature is to the actual feature of the unlearned data. In Table~\ref{tab:metric_stat}, we provide metric statistic summaries for the recovered images in Figure~\ref{fig:cifar10_batch8}. We show that some recovered images with high metric values may still contain identifiable objects. For example, the image (ID 0) of highest LPIPS recovered from the exact unlearning has an LPIPS value of 0.556, which is around the median value as shown in the Figure~\ref{fig:reconst_quality_onesample}, but we can still recognize a contour of an animal (of class ``cat'') from the image.

\begin{table}[t]
\centering
\caption{Metric statistics of recovered images in Figure~\ref{fig:cifar10_batch8}.}
\label{tab:metric_stat}
\resizebox{0.85\linewidth}{!}{
\begin{tabular}{cclcccc}
\toprule
\textbf{\begin{tabular}[c]{@{}c@{}}Unlearning\\ Type\end{tabular}} & \textbf{Metric} & \textbf{Min} & \multicolumn{1}{l}{\textbf{\begin{tabular}[c]{@{}c@{}}Min\\ ID\end{tabular}}} & \textbf{Mean} & \textbf{Max} & \textbf{\begin{tabular}[c]{@{}c@{}}Max\\ ID\end{tabular}} \\ \midrule
\multirow{3}{*}{\begin{tabular}[c]{@{}c@{}}Exact \\ Unlearning\end{tabular}} & MSE & 0.178 & 7 & 0.233 & 0.302 & 4 \\ \cline{2-7} 
 & PNSR & 8.648 & 4 & 13.22 & 25.22 & 5 \\ \cline{2-7} 
 & LPIPS & 0.024 & 5 & 0.339 & 0.556 & 0 \\ \hline
\multirow{3}{*}{\begin{tabular}[c]{@{}c@{}}Approximate\\ Unlearning\end{tabular}} & MSE & 0.055 & 5 & 0.246 & 0.370 & 4 \\ \cline{2-7} 
 & PNSR & 10.40 & 4 & 12.78 & 15.02 & 7 \\ \cline{2-7} 
 & LPIPS & 0.147 & 2 & 0.319 & 0.531 & 6 \\ \bottomrule
\end{tabular}}
\end{table}

\begin{figure}[t]
    \centering
    \includegraphics[width=0.9\linewidth]{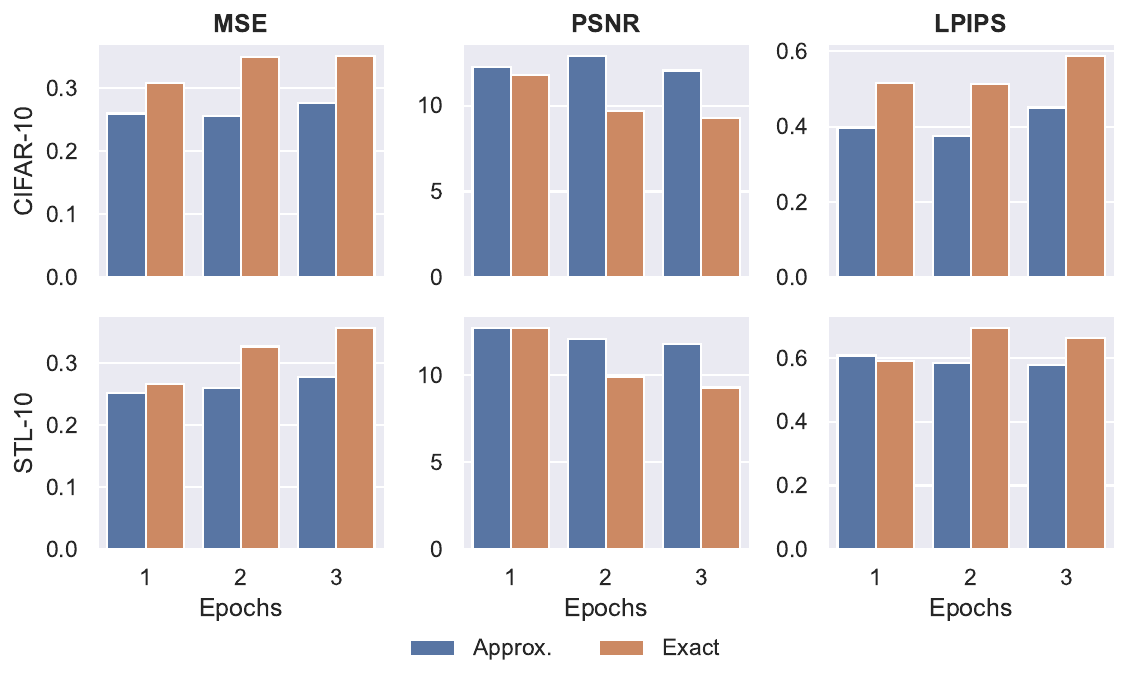}
    \caption{The recover quality for different epochs.}
    \label{fig:unlearn_multi_epoch}
\end{figure}

\noindent\textbf{Fine-tuning Epochs.}
In order to investigate how the fine-tuning epochs on the training dataset affect the feature recovery, we fine-tune the pre-trained model up to three epochs to obtain the final trained model. 
Figure~\ref{fig:unlearn_multi_epoch} presents the quality of the recovered feature for different epochs. 
As the number of epochs increases, the PSNR decreases, and the recovery error measured by MSE and LPIPS slightly augments, especially on exact unlearning. This suggests that more fine-tuning epochs lead to more difficult feature recovery. The reason might be more epochs introduce more randomnesses existing in the training algorithm to the original model and the retrained model, which leads to a biased gradient direction estimation of the unlearned data based on the two model parameter differences.

\begin{figure}[t]
    \centering
    \includegraphics[width=0.95\linewidth]{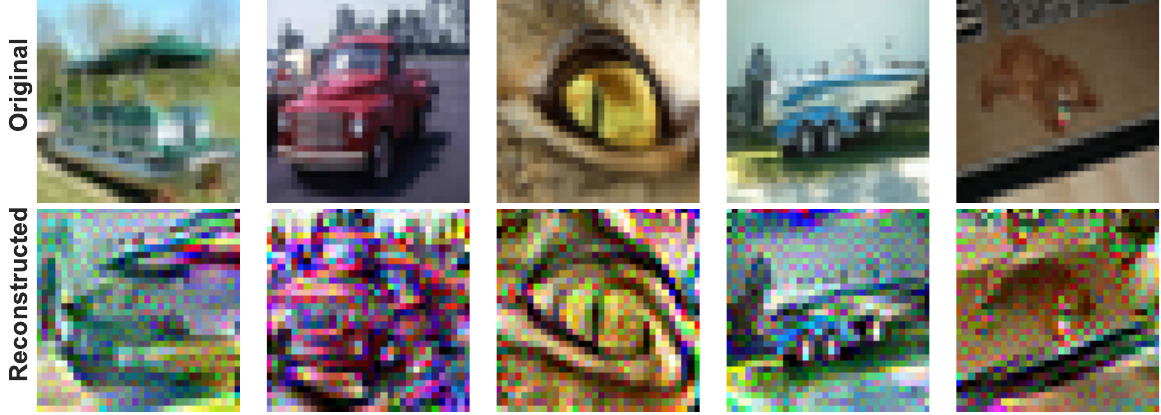}
    \caption{Recovered images in approximate unlearning on CIFAR-10 when unlearning 128 samples at one time.}
    \label{fig:cifar10_multireconst}
\end{figure}

\noindent\textbf{Multiple Unlearning Samples.}
Feature recovery decodes model differences to unlearned sample features via optimization algorithms. Single unlearning inversion is one-to-one, while
multiple unlearning is one-to-many, posing a greater challenge. 
We investigate the possibility of recovering features when multiple samples are unlearned simultaneously. In Figure~\ref{fig:cifar10_multireconst}, we selectively show five readable images recovered from approximate unlearning on CIFAR-10 when unlearning 128 samples at one time. 
As depicted, the objects presented in the recovered images are similar to the objects in the unlearned data. 
Besides, some objects in the recovered images are recognizable, e.g., the cat eye (from class ``cat'') in the middle of Figure~\ref{fig:cifar10_multireconst}. We provide experiments results of all recovered samples for multiple unlearning batches (8, 16, 32) on CIFAR-100 in Figures~\ref{fig::m8},~\ref{fig::m16}, and~\ref{fig::m32} in Appendix~\ref{subsec:additional_experiments}. 
Despite challenges, some features can still be recovered, suggesting that there still is a possibility for correctly recovering the unlearned feature in multiple unlearning.
Furthermore, a larger number of unlearning samples poses greater challenges for unlearning inversion.

\begin{mdframed}[backgroundcolor=white!10,rightline=true,leftline=true,topline=true,bottomline=true,roundcorner=2mm,everyline=true]
\textbf{Takeaway 2.~}
Exact unlearning can also leak feature information of the unlearned data when the training dataset is small. In general, approximate unlearning leaks more feature information than exact unlearning.
\end{mdframed}

\subsection{Ablation Study for Label Inference}
In this subsection, we investigate how different factors affect the performance of label inference. Due to the page limit, we provide the ablation studies of the number of probing samples and type of unlearned class in Appendix~\ref{subsec:additional_experiments_label}.

\begin{figure}[t]
    \centering
    \includegraphics[width=0.9\linewidth]{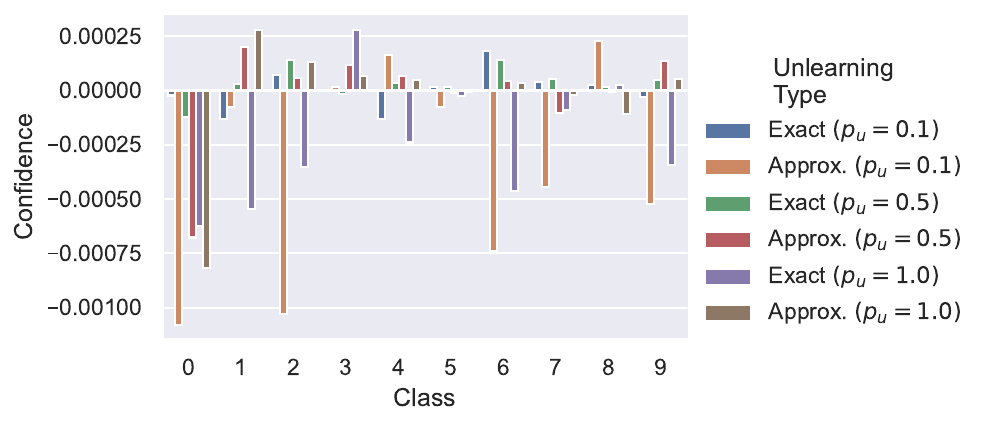}
    \caption{The prediction confidence changes in different unlearning settings. $p_u$ indicates the ratio of unlearned samples in the unlearned class, which is, without loss of generality, set to the class 0. }
    \label{fig:label_infer_oneclass_proportion}
\end{figure}

\noindent\textbf{Proportion of Unlearned Samples in One Class.}
We investigate the label inference effectiveness when unlearning $p_u \in (0,1]$ proportion of samples in a class.
The label of the unlearned samples is set to 0, and the other settings are the same as in Section~\ref{sec:effectiveness}.
Figure~\ref{fig:label_infer_oneclass_proportion} presents the prediction confidence change for $p_u\in\{0.1, 0.5, 1.0\}$. Here, $p_u=1.0$ represents unlearning all samples of class 0, and $p_u=0.1$ represents the case where only 5 samples of class 0 are unlearned.
We found that the label of the unlearned data can be correctly inferred in all cases except for the exact unlearning of $p_u=0.1$. This shows that in general, the label information of the unlearned data can be easily leaked in machine unlearning. Under the case of $p_u=0.1$ in approximate unlearning, we observe that the large prediction confidence drop exists in various unrelated classes, highlighting the difficulty of label inference in extreme cases where a very small number of samples are unlearned.

\begin{table}[t]
\centering
\caption{Inference results of different classes and with less probing samples. The accuracy represents the proportion of the corrected inferred label over the ten classes. The analysis of this table is provided in Appendix~\ref{subsec:additional_experiments_label}.}
\label{tab:diff_class}
\resizebox{0.99\linewidth}{!}{
\begin{tabular}{cccccccccccccc}
\toprule
\multirow{2}{*}{\textbf{\begin{tabular}[c]{@{}c@{}}\# Probing\\ Samples\end{tabular}}} &
  \multirow{2}{*}{\textbf{\begin{tabular}[c]{@{}c@{}}Unlearning\\ Type\end{tabular}}} &
  \multirow{2}{*}{$p_u$} &
  \multicolumn{10}{c}{\textbf{Class ID}} &
  \multirow{2}{*}{\textbf{Acc.}} \\ \cline{4-13}
 &
   &
   &
  \multicolumn{1}{c}{0} &
  \multicolumn{1}{c}{1} &
  \multicolumn{1}{c}{2} &
  \multicolumn{1}{c}{3} &
  \multicolumn{1}{c}{4} &
  \multicolumn{1}{c}{5} &
  \multicolumn{1}{c}{6} &
  \multicolumn{1}{c}{7} &
  \multicolumn{1}{c}{8} &
  9 &
   \\ \midrule
\multirow{6}{*}{\begin{tabular}[c]{@{}c@{}}10 per\\ Class\end{tabular}} &
  \multirow{3}{*}{\begin{tabular}[c]{@{}c@{}}Approximate\\ Unlearning\end{tabular}} &
  0.1 &
  \multicolumn{1}{c}{0} &
  \multicolumn{1}{c}{9} &
  \multicolumn{1}{c}{2} &
  \multicolumn{1}{c}{3} &
  \multicolumn{1}{c}{4} &
  \multicolumn{1}{c}{5} &
  \multicolumn{1}{c}{6} &
  \multicolumn{1}{c}{7} &
  \multicolumn{1}{c}{8} &
  9 &
  0.9 \\ \cline{3-14} 
 &
   &
  0.5 &
  \multicolumn{1}{c}{0} &
  \multicolumn{1}{c}{1} &
  \multicolumn{1}{c}{2} &
  \multicolumn{1}{c}{3} &
  \multicolumn{1}{c}{4} &
  \multicolumn{1}{c}{5} &
  \multicolumn{1}{c}{6} &
  \multicolumn{1}{c}{7} &
  \multicolumn{1}{c}{8} &
  9 &
  1.0 \\ \cline{3-14} 
 &
   &
  1.0 &
  \multicolumn{1}{c}{0} &
  \multicolumn{1}{c}{1} &
  \multicolumn{1}{c}{2} &
  \multicolumn{1}{c}{3} &
  \multicolumn{1}{c}{4} &
  \multicolumn{1}{c}{5} &
  \multicolumn{1}{c}{6} &
  \multicolumn{1}{c}{7} &
  \multicolumn{1}{c}{8} &
  9 &
  1.0 \\ \cline{2-14} 
 &
  \multirow{3}{*}{\begin{tabular}[c]{@{}c@{}}Exact\\ Unlearning\end{tabular}} &
  0.1 &
  \multicolumn{1}{c}{1} &
  \multicolumn{1}{c}{1} &
  \multicolumn{1}{c}{4} &
  \multicolumn{1}{c}{3} &
  \multicolumn{1}{c}{4} &
  \multicolumn{1}{c}{4} &
  \multicolumn{1}{c}{4} &
  \multicolumn{1}{c}{8} &
  \multicolumn{1}{c}{4} &
  4 &
  0.3 \\ \cline{3-14} 
 &
   &
  0.5 &
  \multicolumn{1}{c}{0} &
  \multicolumn{1}{c}{1} &
  \multicolumn{1}{c}{2} &
  \multicolumn{1}{c}{3} &
  \multicolumn{1}{c}{4} &
  \multicolumn{1}{c}{0} &
  \multicolumn{1}{c}{6} &
  \multicolumn{1}{c}{4} &
  \multicolumn{1}{c}{4} &
  9 &
  0.7 \\ \cline{3-14} 
 &
   &
  1.0 &
  \multicolumn{1}{c}{0} &
  \multicolumn{1}{c}{1} &
  \multicolumn{1}{c}{2} &
  \multicolumn{1}{c}{3} &
  \multicolumn{1}{c}{4} &
  \multicolumn{1}{c}{7} &
  \multicolumn{1}{c}{6} &
  \multicolumn{1}{c}{7} &
  \multicolumn{1}{c}{8} &
  9 &
  0.9 \\ \hline
\multirow{6}{*}{\begin{tabular}[c]{@{}c@{}}5 per\\ Class\end{tabular}} &
  \multirow{3}{*}{\begin{tabular}[c]{@{}c@{}}Approximate\\ Unlearning\end{tabular}} &
  0.1 &
  \multicolumn{1}{c}{2} &
  \multicolumn{1}{c}{9} &
  \multicolumn{1}{c}{2} &
  \multicolumn{1}{c}{3} &
  \multicolumn{1}{c}{4} &
  \multicolumn{1}{c}{5} &
  \multicolumn{1}{c}{6} &
  \multicolumn{1}{c}{7} &
  \multicolumn{1}{c}{8} &
  9 &
  0.8 \\ \cline{3-14} 
 &
   &
  0.5 &
  \multicolumn{1}{c}{0} &
  \multicolumn{1}{c}{6} &
  \multicolumn{1}{c}{2} &
  \multicolumn{1}{c}{3} &
  \multicolumn{1}{c}{4} &
  \multicolumn{1}{c}{5} &
  \multicolumn{1}{c}{6} &
  \multicolumn{1}{c}{7} &
  \multicolumn{1}{c}{8} &
  9 &
  0.9 \\ \cline{3-14} 
 &
   &
  1.0 &
  \multicolumn{1}{c}{0} &
  \multicolumn{1}{c}{1} &
  \multicolumn{1}{c}{2} &
  \multicolumn{1}{c}{3} &
  \multicolumn{1}{c}{4} &
  \multicolumn{1}{c}{5} &
  \multicolumn{1}{c}{6} &
  \multicolumn{1}{c}{7} &
  \multicolumn{1}{c}{8} &
  9 &
  1.0 \\ \cline{2-14} 
 &
  \multirow{3}{*}{\begin{tabular}[c]{@{}c@{}}Exact\\ Unlearning\end{tabular}} &
  0.1 &
  \multicolumn{1}{c}{4} &
  \multicolumn{1}{c}{4} &
  \multicolumn{1}{c}{4} &
  \multicolumn{1}{c}{0} &
  \multicolumn{1}{c}{4} &
  \multicolumn{1}{c}{3} &
  \multicolumn{1}{c}{4} &
  \multicolumn{1}{c}{4} &
  \multicolumn{1}{c}{4} &
  4 &
  0.1 \\ \cline{3-14} 
 &
   &
  0.5 &
  \multicolumn{1}{c}{0} &
  \multicolumn{1}{c}{9} &
  \multicolumn{1}{c}{4} &
  \multicolumn{1}{c}{3} &
  \multicolumn{1}{c}{4} &
  \multicolumn{1}{c}{3} &
  \multicolumn{1}{c}{6} &
  \multicolumn{1}{c}{7} &
  \multicolumn{1}{c}{3} &
  9 &
  0.6 \\ \cline{3-14} 
 &
   &
  1.0 &
  \multicolumn{1}{c}{0} &
  \multicolumn{1}{c}{1} &
  \multicolumn{1}{c}{2} &
  \multicolumn{1}{c}{3} &
  \multicolumn{1}{c}{4} &
  \multicolumn{1}{c}{4} &
  \multicolumn{1}{c}{6} &
  \multicolumn{1}{c}{7} &
  \multicolumn{1}{c}{4} &
  9 &
  0.8 \\ \bottomrule
\end{tabular}}
\end{table}

\begin{figure}[t]
    \centering
    \includegraphics[width=0.9\linewidth]{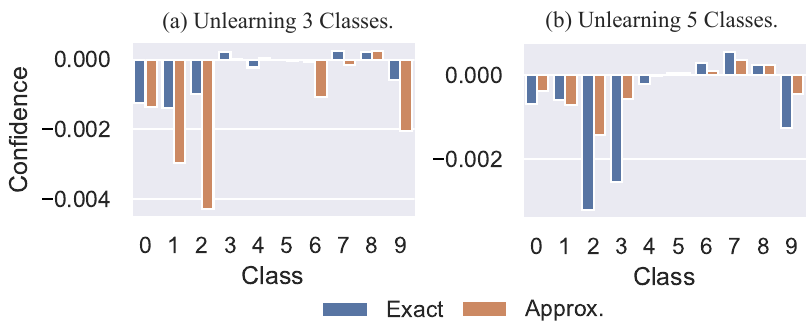}
    \caption{Confidence variation of unlearning data from multiple classes.}
    \label{fig:unlearn_multiclass}
\end{figure}
\noindent\textbf{Multiple Unlearned Classes.}
We evaluate the effectiveness of label inference when unlearning samples from multiple classes.
In our experiments, we consider unlearning samples from 3 classes and 5 classes. We show the prediction confidence change of the probing samples in Figure~\ref{fig:unlearn_multiclass}, where Figure~\ref{fig:unlearn_multiclass} (a) and the Figure~\ref{fig:unlearn_multiclass} (b) demonstrate the cases of the samples from the first three classes (i.e., label 0-2) and five classes (i.e., label 0-4) are unlearned, respectively.
For both cases, the negative confidence change exists in the unlearned classes. However, there are other classes whose confidence values also drop on the probing samples, which is similar to the case of unlearning samples from one class. Based on our top-$K$ strategy for inferring multiple-class described in Section~\ref{sec:attack-method}, the label inference accuracy is $2/3\approx66.7\%$ in the case of three classes and $4/5=80\%$ in the case of five classes. This shows that unlearning inversion attacks for label inference are also effective when samples from multiple classes are unlearned.

\begin{mdframed}[backgroundcolor=white!10,rightline=true,leftline=true,topline=true,bottomline=true,roundcorner=2mm,everyline=true]
\textbf{Takeaway 3.~}
Machine unlearning can leak the label information of the unlearned data in various settings. In general, approximate unlearning leaks more label information than exact unlearning.
\end{mdframed}

\section{Possible Defenses}
To prevent the information leakage of the unlearned data in machine unlearning, we consider the model developer to post-process the unlearned model before uploading it to the server, as the model difference reflects the information of the unlearned data.
Inspired by previous unlearning studies~\cite{wu2020deltagrad,guo2020certified} and the well-known privacy protection method of DP-SGD~\cite{abadi2016deep}, we design three defenses as follows.

\noindent \textbf{Parameter Obfuscation.} Similar to DP-SGD, the developer clips the gradient used for unlearning the samples by $C$ and adds random noise following the Gaussian distribution of variance $\sigma$, which controls the noise magnitude. Note that although clip the gradient may affect the unlearning effectiveness itself, we consider it as an exploration following DP-SGD as a defense to mitigate unlearning inversion attacks.
We set $C=1.2$ following the official tutorial\footnote{\url{https://github.com/pytorch/opacus/blob/main/tutorials/building_image_classifier.ipynb}} of DP-SGD. We add $\sigma\in\{0.001, 0.003, 0.005, 0.007\}$ to achieve different validation accuracy ratios, i.e., the ratio between the validation accuracy of the defended model and that of the undefended unlearned model. Accordingly, we can obtain the validation accuracy ratios of $99.13\%$, $96.37\%$, $90.64\%$, and $84.71\%$, respectively.  

\noindent \textbf{Model Pruning.} The developer deletes a $p$~($0<p<1$) proportion of the least significant parameters in the model using the model pruning technique \cite{zhu2017prune}.
We consider $p\in\{0.7, 0.8, 0.9\}$ to achieve the validation accuracy ratio $95.01\%$, $77.81\%$, and $43.83\%$.

\noindent \textbf{Fine-tuning.} The developer can fine-tune the unlearned model on additional samples so that the model difference becomes less accurate for unlearning inversion. We consider the developer fine-tunes the unlearned model on an additional dataset of 100 samples, varying fine-tuning rates (\textit{lr}) in \{0.0001, 0.0005, 0.001, 0.005, 0.01\}. Accordingly, the unlearned model has the validation accuracy ratios in [35\%, 135\%]. Note that the accuracy ratio can be above 100\% because fine-tuning may improve the model performance.

\begin{figure}[t]
    \centering
    \includegraphics[width=0.9\linewidth]{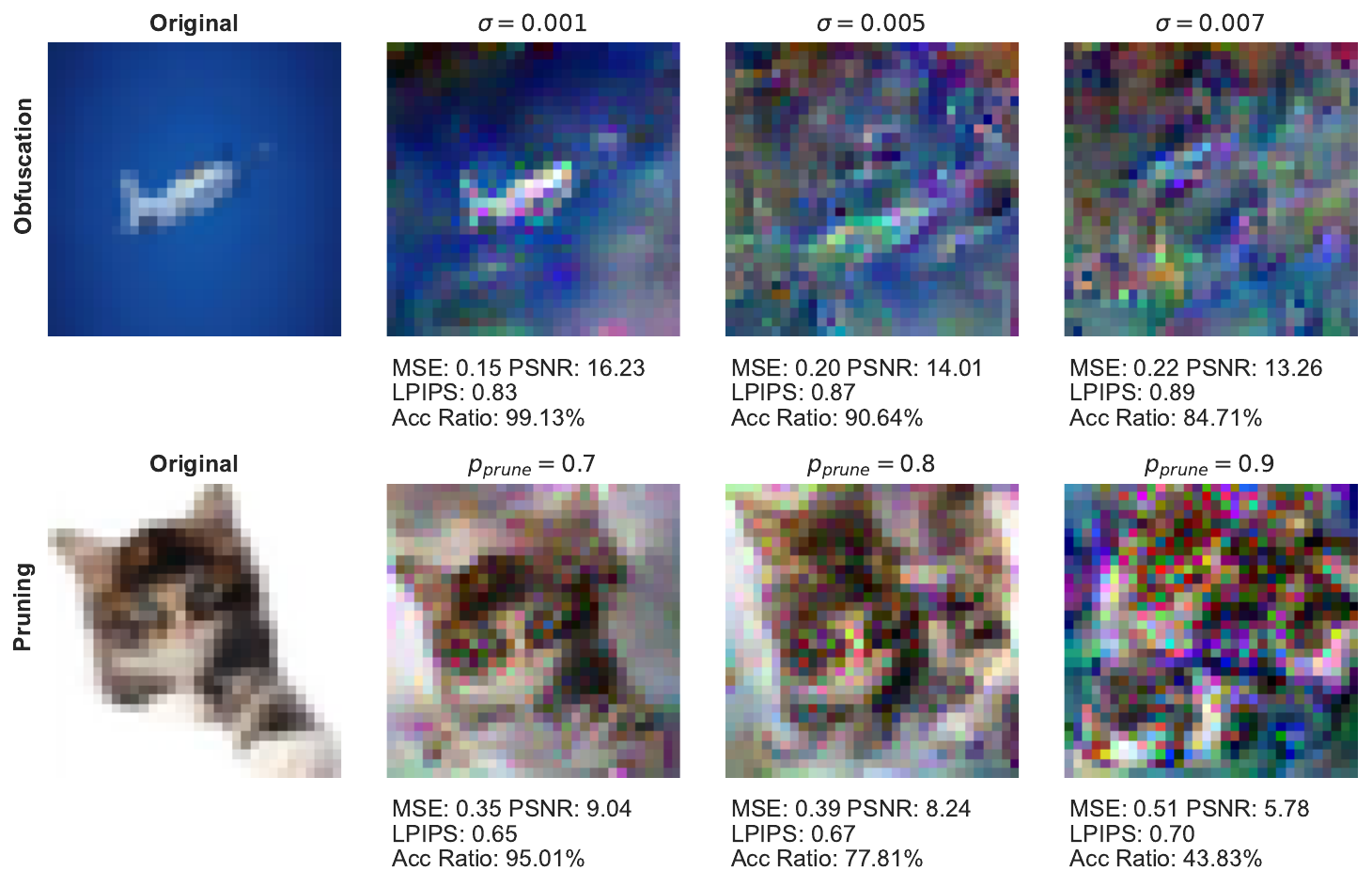}
    \caption{Effectiveness of parameter obfuscation and model pruning in defending against unlearning inversion attacks for feature recovery. }
    \label{fig:defense_feature_example}
\end{figure}

\begin{figure}[t]
    \centering
    \includegraphics[width=0.9\linewidth]{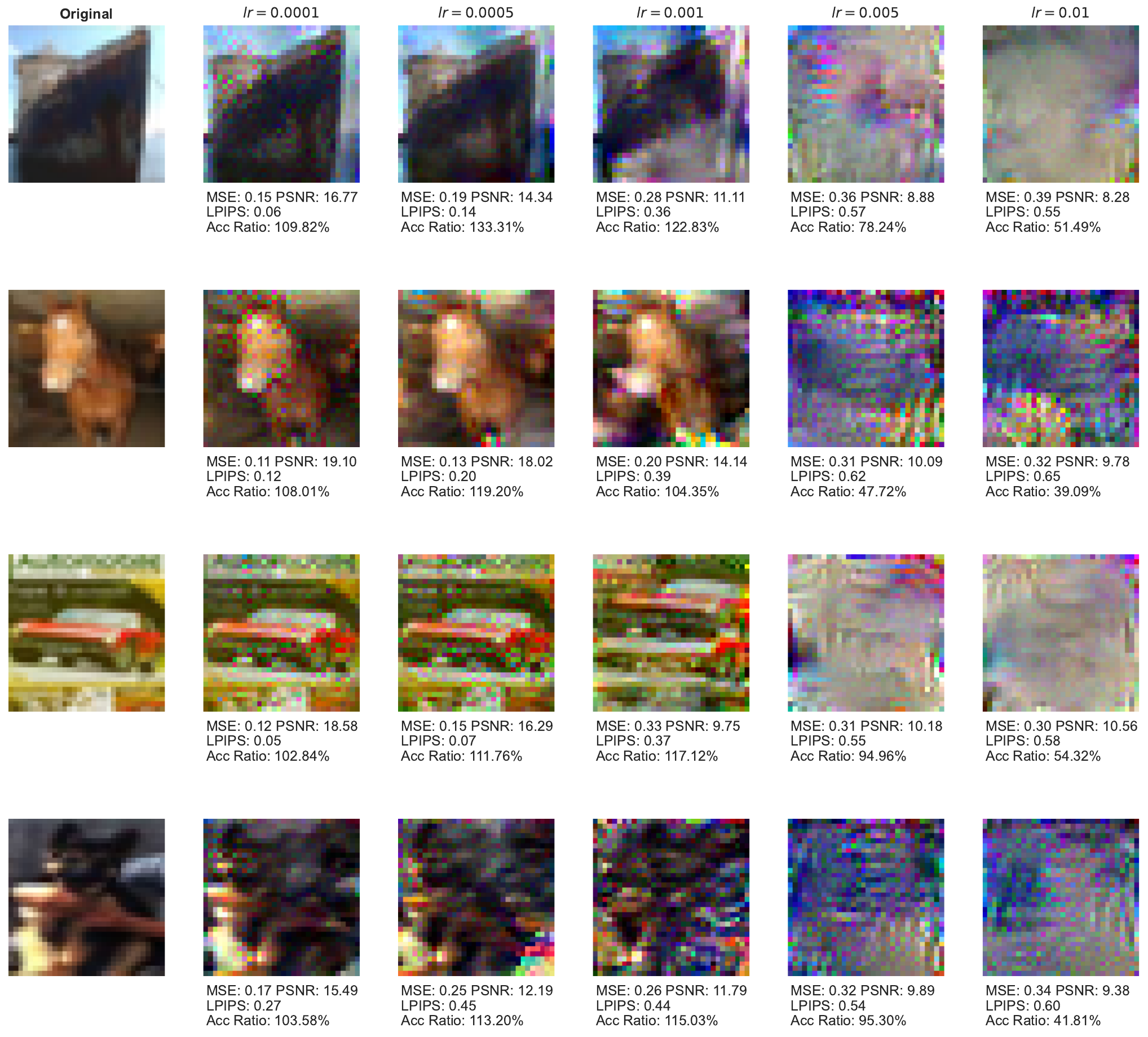}
    \caption{Effectiveness of fine-tuning in defending against unlearning inversion attacks for feature recovery.}
    \label{fig::ft}
\end{figure}

\begin{table}[!t]
    \centering
    \caption{Trade-off between model utility (measured by averaged accuracy ratio) and privacy (measured by defense parameter).}
    
    \resizebox{0.8\linewidth}{!}{
    \begin{tabular}{llccc}
\toprule
\textbf{Defense} & \begin{tabular}[c]{@{}c@{}}\textbf{Defense} \\ \textbf{Parameter} \end{tabular} & \begin{tabular}[c]{@{}c@{}}\textbf{Parameter} \\ \textbf{Value} \end{tabular}  & \textbf{LPIPS} & \begin{tabular}[c]{@{}c@{}c@{}}\textbf{Averaged} \\ \textbf{Accuracy} \\ \textbf{Ratio} \end{tabular} 
\\ \midrule
\multirow{4}{*}{\begin{tabular}[c]{@{}c@{}}Parameter \\obfuscation\end{tabular}} & \multirow{4}{*}{\begin{tabular}[c]{@{}c@{}} $\sigma$\end{tabular}} & 0.001 & 0.45 & 0.99 \\ \cline{3-5} 
 & & 0.003 & 0.47 & 0.97 \\ \cline{3-5} 
 & & 0.005 & 0.49 & 0.93 \\ \cline{3-5} 
 & & 0.007 & 0.47 & 0.86 \\ \hline
\multirow{3}{*}{\begin{tabular}[c]{@{}c@{}}Model \\pruning\end{tabular}} & \multirow{3}{*}{\begin{tabular}[c]{@{}c@{}} $p$ \end{tabular}} & 0.7 & 0.45 & 0.96 \\ \cline{3-5} 
 & & 0.8 & 0.48 & 0.83 \\ \cline{3-5} 
 & & 0.9 & 0.56 & 0.59 \\ \hline
\multirow{5}{*}{\begin{tabular}[c]{@{}c@{}}Fine-tuning\end{tabular}} & \multirow{5}{*}{\begin{tabular}[c]{@{}c@{}} $lr$ \end{tabular}} & 0.0001 & 0.43 & 1.04 \\ \cline{3-5} 
 & &  0.0005 & 0.45 & 1.12 \\ \cline{3-5} 
 & & 0.001 & 0.50 & 1.17 \\ \cline{3-5} 
 & & 0.005 & 0.56 & 1.06 \\ \cline{3-5} 
 & & 0.01 & 0.57 & 0.71 \\ \bottomrule
\end{tabular}}
    
    \label{table::defense}
  
\end{table}

We evaluate the three defenses on CIFAR-10 for feature recovery in approximate unlearning under single unlearning settings, which in general leak more information than other settings. Figure~\ref{fig:defense_feature_example} and~\ref{fig::ft} exhibits the recovered examples under different defense intensities. We observe that larger defense parameters $\sigma$, $p$, or \textit{lr} can worsen the quality of the recovered features but also scarify the unlearned model's utility.
For example, on the first row in Figure~\ref{fig:defense_feature_example}, the recovered image becomes blurring for $\sigma \geq 0.005$, which degrades the model's validation accuracy by 93.14\%.

\noindent \textbf{Privacy and Utility Trade-off.} Three defenses—parameter obfuscation, model pruning, and fine-tuning balance privacy and utility trade-offs. We offer CIFAR-10 results in Table~\ref{table::defense} to
guide practitioners in selecting defense parameters. This table is obtained by repeating 100 times experiments on feature recovery by randomly selecting 100 unlearned samples, which showcases defenses that effectively mitigate unlearning inversion with minimal impact on model utility, as reflected by the accuracy ratio. Using an LPIPS threshold of 0.5 (the inverted feature becomes blurring) for securing from unlearning inversion, a lower LPIPS value indicates increased difficulty for unlearning inversion. Table~\ref{table::defense} reveals optimal parameter values: 0.005 for parameter obfuscation, [0.8, 0.9] for model pruning, and 0.001 for fine-tuning. While defense parameters can vary across applications, observing when unlearning inversion becomes meaningless enables finding optimal parameter settings of different defense methods in different applications.

\begin{mdframed}[backgroundcolor=white!10,rightline=true,leftline=true,topline=true,bottomline=true,roundcorner=2mm,everyline=true]
\textbf{Takeaway 4.~}
Parameter obfuscation, model pruning, and fine-tuning are effective in defending against unlearning inversion attacks, while at a cost of heavily sacrificing the unlearned model's utility.
\end{mdframed}

\section{Limitations of the Attacks}
\noindent \textit{Multiple Unlearning.} In multiple unlearning cases, it may not be possible to recover each feature of unlearned samples, as demonstrated in Figure~\ref{fig::m8},~\ref{fig::m16}, and~\ref{fig::m32}. This is due to the limitation of the optimization based feature recovery method: multiple unlearned samples can have different possible permutations and thus make the optimizer
difficult to find correct gradients. However, we do observe that some recovered features are recognizable. If such recovered features contain highly sensitive information, privacy can be broken in machine unlearning in multiple unlearning scenarios.

\noindent \textit{Label Information in Feature Recovery.} In the experiments of feature recovery, we assume the label information is known for expediting optimization. Feature recovery can also be feasible by setting the label as an unknown variable in the optimization process, with higher computational costs. Results in Figure~\ref{fig:label-unknown} show that similar attack performance can be obtained in label known and unknown settings.

\noindent \textit{User Attack.} In this paper, we consider a user, with only black-box access to the model, who attempts to invert the label of an unlearned sample. There is a possibility for the user to recover the feature: \textit{i}) employing model inversion attacks~\cite{fredrikson2015model}, which may only yield representative features of a class instead of the exact feature of the unlearned data; \textit{ii}) training two surrogate models using model extraction attacks~\cite{tramer2016stealing} based on access to the original and unlearned models and then using the model difference for feature inversion. However, this approach's effectiveness may be compromised due to the inaccurate mapping of the model difference to the feature.

\section{Discussion}\label{sec:discussion}
\noindent \textbf{Trade-offs in Machine Unlearning.} In existing studies of machine unlearning, two aspects are usually considered in the trade-offs in exact unlearning and approximate unlearning: effectiveness and efficiency. Exact unlearning has a perfect unlearning effectiveness guarantee compared to approximate unlearning while at the cost of efficiency. The proposed unlearning inversion attacks raise a new aspect that has to be taken into consideration, i.e., information leakage. As depicted in the experiments, approximate unlearning leaks more information about the unlearned data than exact unlearning, while the former is much more efficient than the latter. Our study highlights the need for careful consideration in achieving a new balance among unlearning effectiveness, efficiency, and information leakage.

\noindent \textbf{Defense against Label Inference.} We discuss two potential defenses for mitigating unlearning inversion attacks for feature recovery, as the feature of an unlearned sample usually contains more information than the label. To defend against unlearning inversion attacks for label inference, one promising way from the server's perspective is to provide only the predicted label of a query sample instead of the full confidence scores of all classes, which are leveraged by the user for label inference. This also preserves the utility of the unlearned model, as the model parameters are not altered.
However, as shown in studies of membership inference attacks \cite{li2021membership,choquette2021label}, providing the label is not safe to fully mitigate the attack because of more advanced label-only membership inference attacks. Similarly, providing only the prediction label may also not be safe to defend against unlearning inversion attacks, as attacks and defenses are like arms races, and more advanced unlearning inversion attacks might be proposed.

\noindent\textbf{Privacy Disparity.} As shown in the experiments, we find that some unlearned samples' features and label information are more difficult to recover and infer than others, indicating that different samples can suffer different privacy vulnerabilities in machine unlearning. This phenomenon aligns with the findings in existing studies~\cite{song2021systematic,long2020pragmatic,zhong2022understanding,carlini2022membership}, where they demonstrate disparate vulnerabilities of membership privacy risks of the training samples that exist on machine learning models. The disparate vulnerability of the unlearned samples may motivate future studies to design sample-specific unlearning methods to enhance the privacy protection of more easily attacked samples.

\noindent \textbf{Measurement Tool.} Although we investigate unlearning inversion attacks from the perspective of revealing the information of the unlearned data, they can be leveraged as a measurement tool for measuring the information leakage in machine unlearning based on the attack performance. In particular, they enable us to measure other unlearning methods not evaluated in this paper. In addition, as a measurement tool, unlearning inversion attacks inspire future research on machine unlearning and promote the development of safer unlearning methods without severely leaking the information of the unlearned data.

\section{Conclusion}
In this paper, we perform the first investigation to study to what extent the machine unlearning methods can leak the information of the unlearned data. Concretely, we propose unlearning inversion attacks that, given only access to the original and unlearned models, can reveal the feature and label information of the unlearned data. Extensive evaluations demonstrate unlearning inversion attacks can effectively reveal the information on both exact and approximate unlearning methods. To mitigate the attacks, we identify two possible defenses, i.e., parameter obfuscation and model pruning. We evaluate the two defenses and find that they can mitigate the attack while at the cost of reducing the utility of the unlearned model. As machine unlearning becomes increasingly important, the findings in this paper serve as an initiation exploration in understanding the privacy vulnerability in current machine unlearning techniques, emphasizing the importance of avoiding information leakage during the machine unlearning process and inspiring future research to develop more secure unlearning mechanisms.

\section*{Acknowledgments}

This work is supported in part by Australian Research Council (ARC) DP240103068. Minhui Xue is supported by CSIRO – National Science Foundation (US) AI Research Collaboration Program. Minhui Xue and Shuo Wang are the corresponding authors of this paper.

\bibliographystyle{IEEEtran}
\bibliography{reference}

\appendices

\section{Experimental Settings}\label{appendix:settigns}
\noindent \textbf{Dataset.} We provide the detailed description of the datasets used in the experiments as follows. Since CIFAR-10, CIFAR-100, and STL-10 are widely used in image classification tasks, we do not explicitly introduce them here due to page limits.
\begin{itemize}[leftmargin=*]
\item \textbf{Chest X-Rays.}
It compromised by 5,216 training images and 624 images for test.
The images are single-channel and have dimension $128\times128$.
The data are categorized into two classes: normal and pneumonia. 
\item \textbf{Germany Credit.}
This is a binary classification dataset consisting of 1,000 samples in total.
We randomly choose 80\% of the whole dataset for training and leave the rest for test.
The tabular data are of dimension 24 and we apply the max-min normalization (i.e., scale data into $[0,1]$ for each feature) to stabilize training.
\end{itemize}

\noindent \textbf{Metric.} We provide the detailed description of the MSE, PSNR, and LPIPS metrics as follows.
\begin{itemize}[leftmargin=*]
    \item \textit{MSE}: for the ground truth feature $\bm{x}$ and the recovered feature $\bm{x}'$, we compute the Mean Squared Error (MSE) $\sqrt{||\bm{x}-\bm{x}
'||_2^2 / \text{dim}(\bm{x})}$, where $\text{dim}(\bm{x})$ is the dimension.
    \item \textit{PSNR}: We use another common metric, the peak-to-noise ratio (PNSR) between the original and the recovered features for measuring the quality of the recovered feature.
    \item \textit{LPIPS}~\cite{DBLP:conf/cvpr/ZhangIESW18}: We use LPIPS to measure the perceptual similarity between the original and recovered features following prior work~\cite{yue2023gradient,yin2021see}, as it aligns well with human perception.
    We use VGG-16~\cite{simonyan2014very} as backbone model to compute LPIPS.
    Lower LPIPS indicates higher perceptual similarity. 
\end{itemize}

\noindent \textbf{Machine Unlearning Methods.} We detail the two machine unlearning methods evaluated in the experiments as follows:
\begin{itemize}[leftmargin=*]
\item \textbf{Exact Unlearning.} We select retraining as the exact unlearning method in our experiments. Retraining serves as an undisputed baseline in exact machine unlearning because the resulting unlearned model is not trained on the unlearned sample. Formally, let $\bm{x} \in \mathcal{D}_u$ be the sample that is required to be removed. In our experiments, the unlearned model $M_u$ is obtained by fine-tuning $M_0$ on the remaining $\mathcal{D}_u / \bm{x}$ for the same epoch as the original model $M$ trained on $\mathcal{D}_u$.

\item \textbf{Approximate Unlearning.} We select the single gradient unlearning method in~\cite{thudi2022unrolling} as the approximate unlearning method because of its advantage of efficiency. There are other approximate unlearning methods~\cite{wu2020deltagrad,guo2020certified} based on the Newton's method~\cite{byrd1994representations}. However, they usually require to compute the Hessian matrix or its reverse of the model, which is difficult to be calculated for DNNs. In the MLaaS settings, the model developer is more likely to leverage a light-weight approximate unlearning method for machine unlearning as it can save the time and computational cost. The single gradient unlearning method only depends on the initial weights of the pre-trained model $M_0$ and is inexpensive to obtain the unlearned model by directly adding back the gradient of the unlearned sample to the original trained model. Formally, the unlearned model $M_u$ is obtained by:
\begin{equation}
    M_u=M+\nabla_u,
\end{equation}
where
\begin{equation}
    \nabla_u=\frac{\eta m}{b}\frac{\partial \mathcal{L}\left(f_{M_0}(\bm{x}), {y}\right)}{\partial M_0},
\end{equation}
where $(\bm{x},y)$ is the unlearned sample, $\eta$ is the learning rate in the fine-tune process, $b$ is the fine-tune batch size, and $m$ is the number of fine-tune epochs for unlearning. Note that in unlearning inversion attacks for feature recovery, the unlearning parameters are not available for the server. This means the server can only obtain $\nabla_u$ but can not obtain the precise gradient $\frac{\partial \mathcal{L}\left(f_{M_0}(\bm{x}), {y}\right)}{\partial M_0}$ of the unlearned sample, which is a challenge for the feature recovery.

\end{itemize}

\begin{figure}[t]
    \centering
    \includegraphics[width=0.65\linewidth]{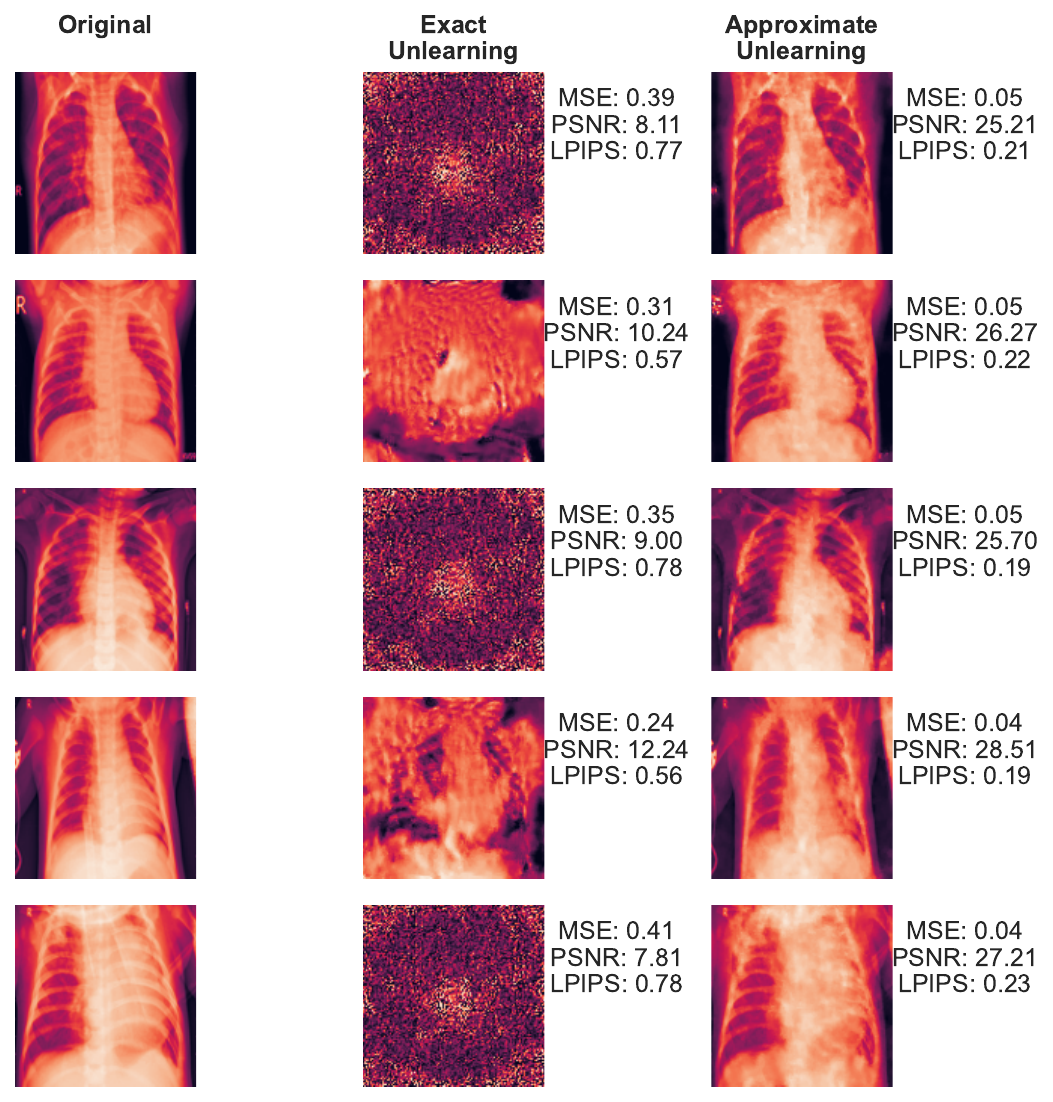}
    \caption{Feature recovery on Chest X-Rays dataset.}
    \label{fig::xray}
\end{figure}

\begin{figure}[t]
    \centering
    \includegraphics[width=0.8\linewidth]{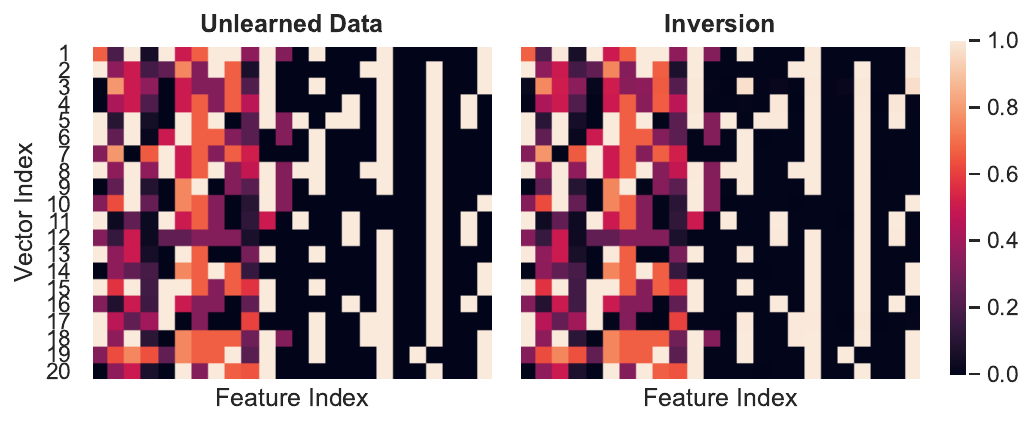}
    \caption{Feature recovery on German Credit dataset. Each line represents a vector.}
    \label{fig::credit}
\end{figure}

\begin{figure}[t!]
    \centering
  \subfloat[CIFAR-10]{%
       \includegraphics[width=0.48\linewidth]{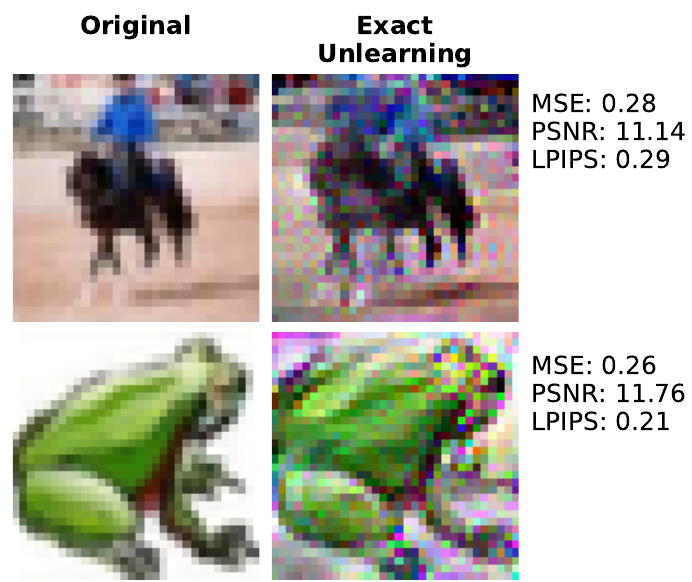}}
  \subfloat[STL-10]{%
        \includegraphics[width=0.46\linewidth]{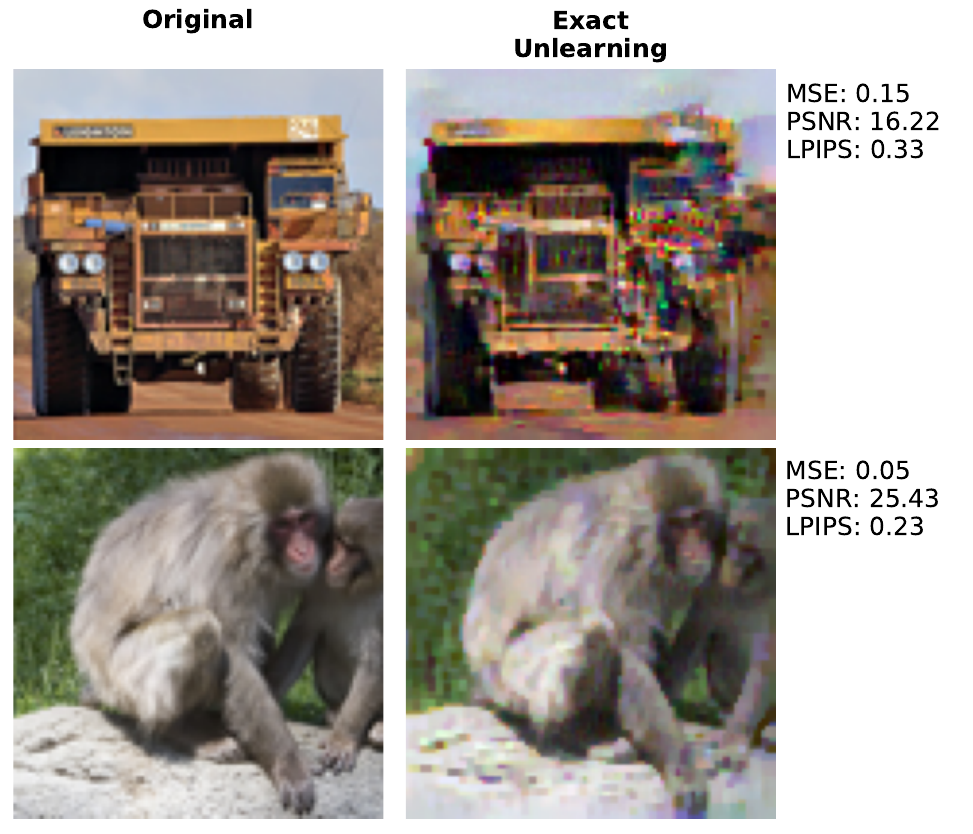}}
  \caption{The recovered features in exact unlearning on CIFAR-10 (left) and STL-10 (right) with a training dataset size of 2.}
  \label{fig:effective-single-unlearn}
\end{figure}

\section{Additional Experiments for Feature Recovery}
\label{subsec:additional_experiments}
\noindent \textbf{Feature Recovery on Additional Datasets.} We provide additional experiments of feature recovery under single unlearning settings on Chest X-Rays and Germany Credit datasets, which help to understand how machine unlearning may leak private information across different domains. 
Results in Figure~\ref{fig::xray},~\ref{fig::credit} reveal high fidelity in recovered X-ray images and accurate retrieval of sensitive numerical data.
These findings confirm the privacy leakage of machine unlearning can indeed happen in sensitive domains and highlight the need for thorough assessments of privacy risks associated with machine unlearning in practical applications.

\noindent\textbf{Feature Recovery in Exact Unlearning.} 
We provide an extreme case of feature recovery in exact unlearning, where the training dataset consists of only two samples, and the original model and the unlearned model differ from one sample. The experimental results serve as an exploration of the information leakage in exact unlearning under extreme cases.
Figure~\ref{fig:effective-single-unlearn} visualize the recovered features of the unlearned samples in CIFAR-10 and STL-10.
In both cases, the left column forms the original samples, and the recovered samples are on the right column.
As we can see, exact unlearning can also leak feature information of the unlearned data when the training dataset is very small.

\begin{figure}[!t]
    \centering
    \includegraphics[width=0.92\linewidth]{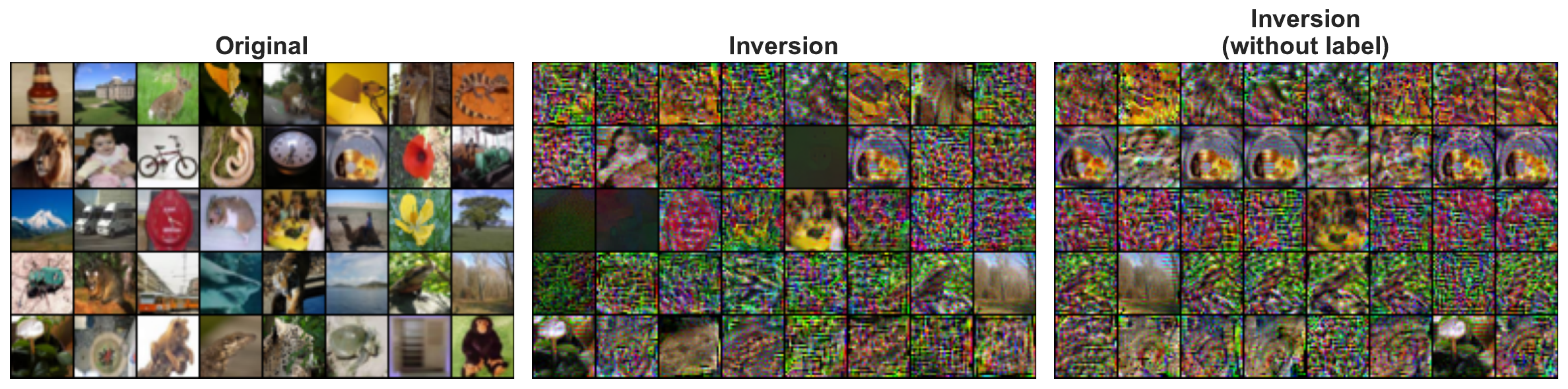}
    \caption{Comparison between feature inversion with (middle) and without (right) label information on CIFAR-100 in the multiple unlearning case. The unlearning and recovery batch size is 8 (i.e., we unlearn and recover 8 random samples). Each row represents a different unlearning batch.}
    \label{fig:label-unknown}
\end{figure}

\noindent\textbf{Feature Recovery with Label as a Variable.} We explore whether unlearning inversion attacks for feature recovery are feasible when considering the label as an unknown variable in the optimization process. Here, we use a multiple unlearning case of 8 samples on CIFAR-100 as a demonstration. We initialize the label as a uniform prediction vector and optimize the label together with the feature to be recovered.
In Figure~\ref{fig:label-unknown}, we present examples recovered from approximate unlearning. As depicted, for the recovered samples with recognizable objects, we can observe a similar attack performance in label known and label unknown cases, indicating that feature inversion attacks can also work when setting the label as an unknown variable.

\begin{figure}[!t]
    \centering
    \includegraphics[width=0.9\linewidth]{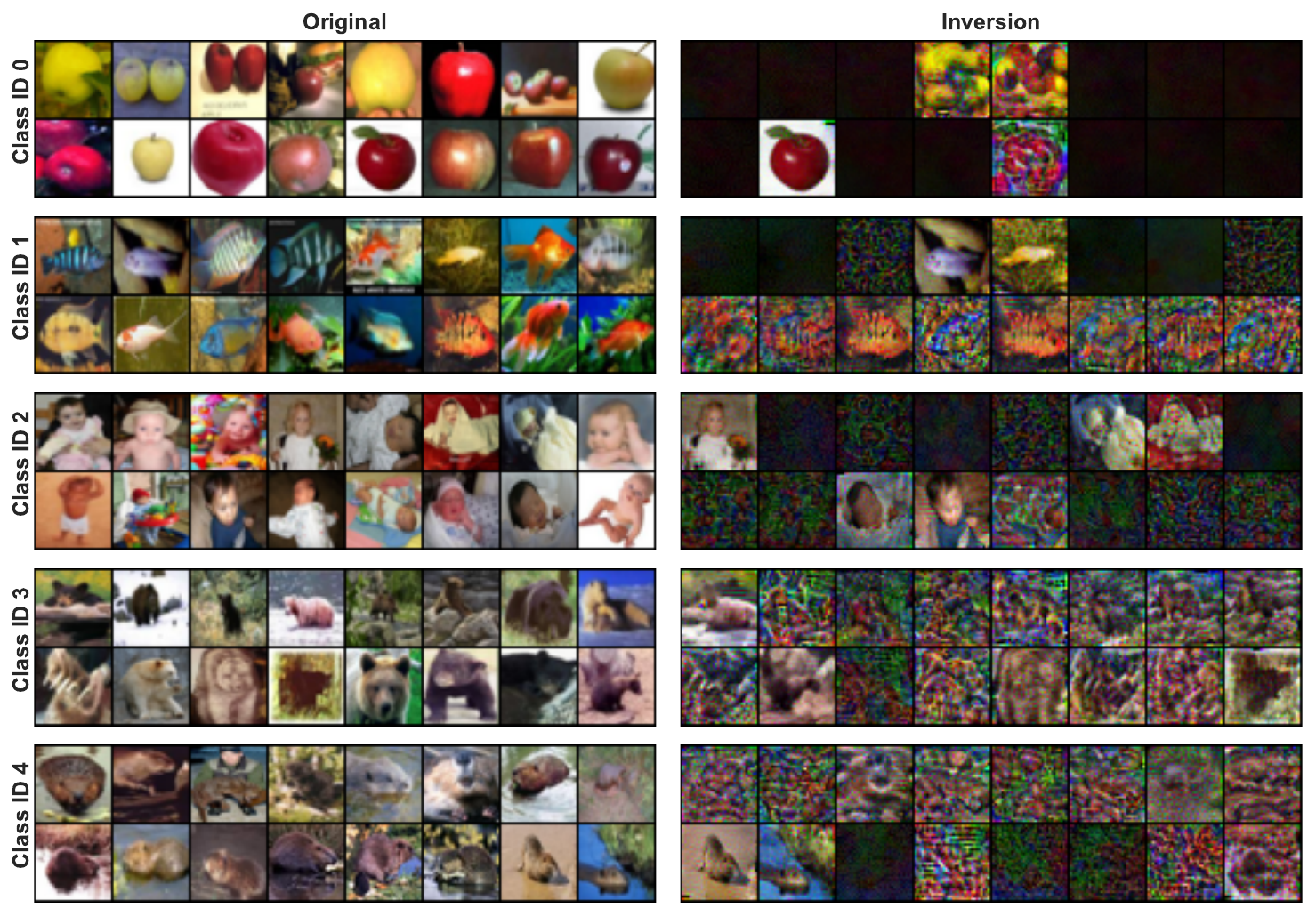}
    \caption{Feature inversion of multiple samples (8 samples) within a single class in CIFAR-100. Each row represents a different random unlearning batch. We choose the first 5 classes of CIFAR-100.}
    \label{fig::class}
\end{figure}

\begin{figure}[!t]
    \centering
    \includegraphics[width=0.9\linewidth]{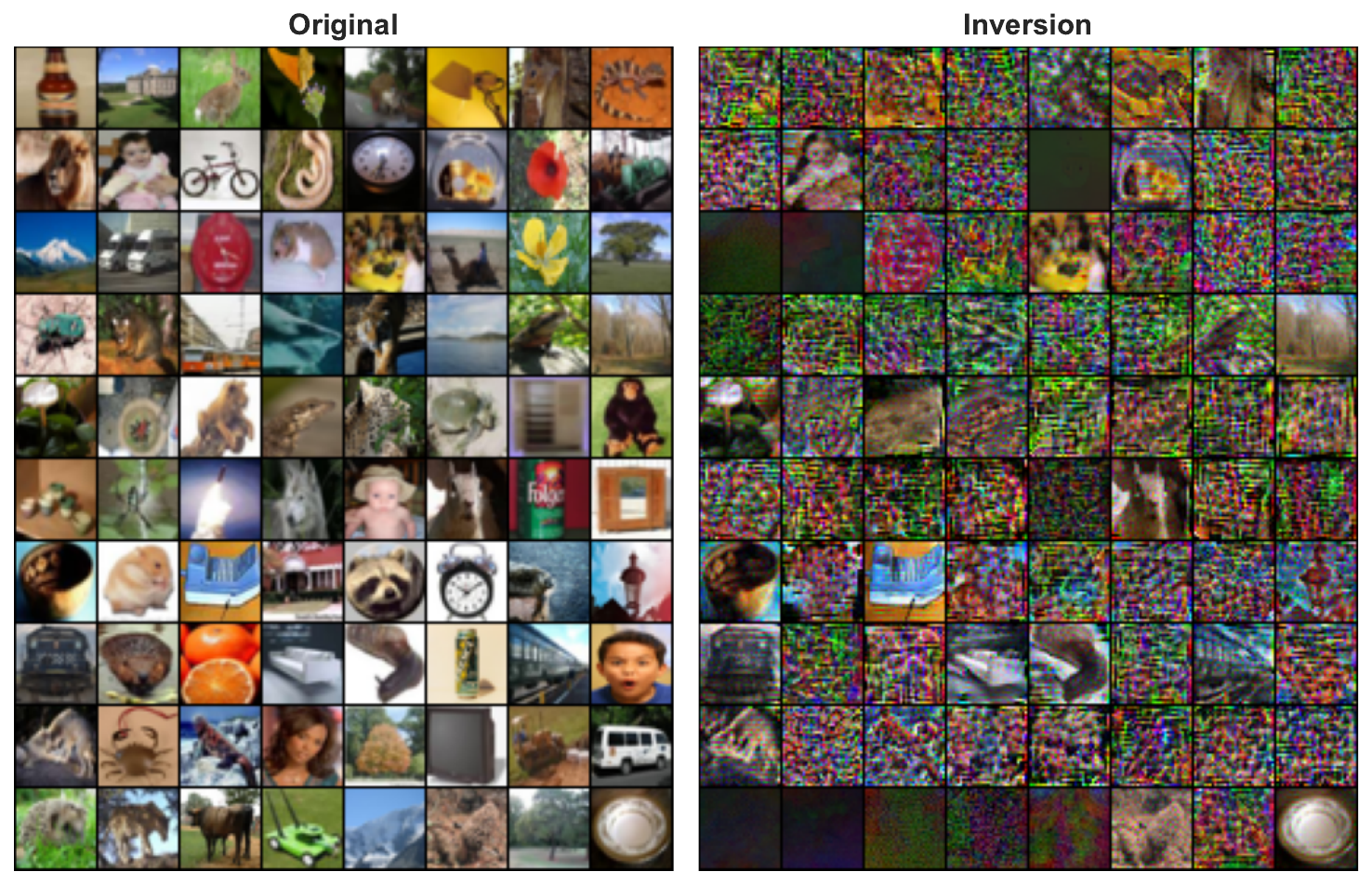}
    \caption{Feature recovery in multiple unlearning of size 8 on CIFAR-100. Each row represents a different random unlearning batch.}
    \label{fig::m8}
\end{figure}

\begin{figure}[!t]
    \centering
    \includegraphics[width=1.0\linewidth]{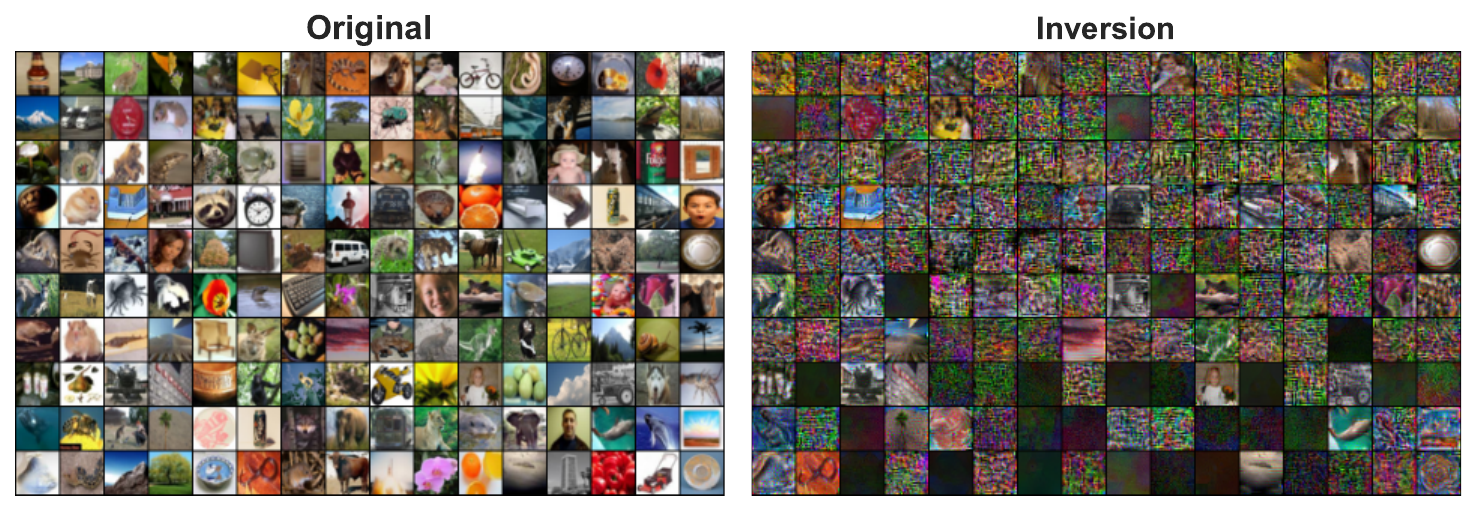}
    \caption{Feature recovery in multiple unlearning of size 16 on CIFAR-100. Each row represents a different random unlearning batch.}
    \label{fig::m16}
\end{figure}

\begin{figure}[!t]
    \centering
    \includegraphics[width=1\linewidth]{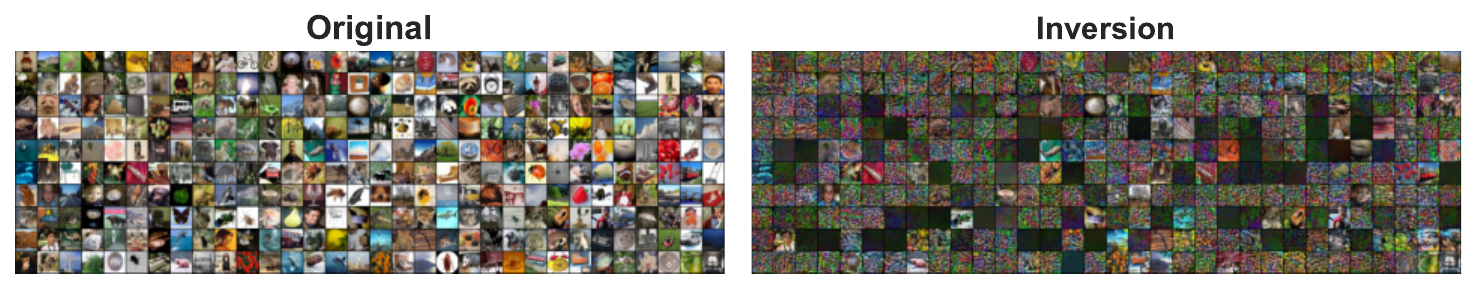}
    \caption{Feature recovery in multiple unlearning of size 32 on CIFAR-100. Each row represents a different random unlearning batch.}
    \label{fig::m32}
\end{figure}

\noindent \textbf{Feature Recovery in Multiple Unlearning.} We provide two additional experiments of feature recovery in multiple unlearning cases on CIFAR-100 when unlearned models are obtained by approximate unlearning. The first experiment in Figure~\ref{fig::class} studies multiple unlearning samples from the same class. The second experiment investigates multiple unlearning of 8, 16, and 32 samples randomly selected from different classes in Figure~\ref{fig::m8},~\ref{fig::m16}, and~\ref{fig::m32}. Results show that feature recovery in multiple unlearning cases is indeed difficult, as not many samples can be recovered with high quality. However, we do observe that some samples are recovered well, showing that multiple unlearning still has the potential to leak the feature information.

\vspace*{15px}

\section{Additional Ablation Study for Label Inference}
\label{subsec:additional_experiments_label}

\noindent\textbf{Number of Probing Samples.}  We investigate whether an attaker can leverage fewer probing samples than 10 to perform label inference, as fewer probing samples can save the attack effort and computation cost.
The lower part of Table~\ref{tab:diff_class} lists the inference accuracy when the attacker uses 5 probing samples.
Although we only use 5 probing samples per class (50\% lower than the default settings of 10 probing samples per class) to infer the label of the unlearned data, the inference accuracy does not significantly decrease.
Compared to the default setting, the accuracy drops at most 0.2 in exact unlearning with $p_u=0.1$ and remains unchanged for approximate unlearning with $p_u=1.0$.
This reflects that our label inference method in unlearning inversion attacks can still be effective when the attacker has limited resources to create plenty of probing samples.

\noindent\textbf{Type of the Unlearned Data Class.}
Next, we investigate whether the label of the samples other than class 0 can be correctly inferred when such samples are unlearned.
Table~\ref{tab:diff_class} (upper part) presents the inference results for ten classes in exact unlearning and approximate unlearning with $p_u\in\{0.1, 0.5, 1.0\}$.
The unlearned label is predicted by selecting the class of the probing samples that has the largest confidence drop.
As we can see, approximate learning leaks more label information than exact unlearning because the inference accuracy is close to 1.
On the other hand, there are classes easier to be leaked than the others.
For example, classes 3 and 4 can be correctly predicted even when only 5 samples are unlearned in exact unlearning.

\clearpage
\section{Meta-Review}

The following meta-review was prepared by the program committee for the 2024
IEEE Symposium on Security and Privacy (S\&P) as part of the review process as
detailed in the call for papers.

\subsection{Summary}
This paper explores the privacy vulnerabilities in machine unlearning processes, particularly focusing on how machine unlearning can inadvertently leak sensitive information about unlearned data. The authors introduce unlearning inversion attacks, which can reveal both feature (white-box) and label (black-box) information of unlearned samples by accessing the original and unlearned models.

\subsection{Scientific Contributions}
\begin{itemize}
\item Identifies an Impactful Vulnerability.
\item Provides a Valuable Step Forward in an Established Field.
\item Creates a New Tool to Enable Future Science.

\end{itemize}

\subsection{Reasons for Acceptance}

This paper provides a valuable step forward in the literature of machine unlearning, which is an important research direction. The authors uncover a significant oversight in current unlearning practices by demonstrating how the adversary can exploit the differences between original and unlearned models to infer sensitive information about the unlearned data.

\end{document}